\documentclass[sigconf, authordraft, review = false]{acmart}
\AtBeginDocument{%
  }

\acmDOI{}
\acmISBN{}
\acmYear{}
\setcopyright{none}
\acmConference[]{}{}{}
\usepackage[yyyymmdd]{datetime}

\usepackage{tabularx}
\usepackage{dcolumn} %Aligning numbers by decimal points in table columns
\newcolumntype{d}[1]{D{.}{.}{#1}}

\usepackage{subfig}
\usepackage[capitalize]{cleveref}
\usepackage[dvipsnames]{xcolor}
\usepackage[most]{tcolorbox}

\usepackage{booktabs}
\usepackage{array}
\usepackage{subcaption}
\usepackage{enumitem}
\usepackage{listings}
\usepackage{makecell}
\usepackage{adjustbox}
\usepackage{tikz}
\usepackage{xcolor}
\usepackage{float}

\newcommand{\blackcircled}[1]{%
  \tikz[baseline=(char.base)]{
    \node[circle,fill=black,text=white,inner sep=.5pt] (char) {\textbf{\sffamily #1}};
  }%
}
\usepackage{todonotes}
\let\xtodo\todo
\renewcommand{\todo}[1]{\xtodo[inline,color=green!50]{#1}}

\begin{document}

% \title[Conversations in Space]{Conversations in Space: Structuring Non-Linear LLM Interactions on a Canvas}

% \title[]{What Does the Branch Know? A Five-Day Field Study of Non-Linear LLM Conversation}

\title{Conversations in Space: Non-Linear LLM Interaction in Everyday Use}

\author{Rifat Mehreen Amin}
\orcid{0000-0003-4279-7778}
\affiliation{%
  \institution{LMU Munich}
  \city{Munich}
  \country{Germany}
}
\email{rifat.amin@ifi.lmu.de}

\author{Alperen Adatepe}
\orcid{0009-0005-0924-0347}
\affiliation{%
  \institution{LMU Munich}
  \city{Munich}
  \country{Germany}
}
\email{adatepe.alperen@campus.lmu.de}

\author{Daniela Fernandes}
\orcid{0009-0006-1332-7485}
\affiliation{%
  \institution{Aalto University}
  \city{Espoo}
  \country{Finland}
}
\email{daniela.dasilvafernandes@aalto.fi}

\author{Daniel Buschek}
\orcid{0000-0002-0013-715X}
\affiliation{%
 \institution{University of Bayreuth}
 % \department{Department of Computer Science}
  \city{Bayreuth}
  \country{Germany}}
\email{daniel.buschek@uni-bayreuth.de}

\author{Andreas Butz}
\orcid{0000-0002-9007-9888}
\affiliation{%
 \institution{LMU Munich}
  \city{Munich}
  \country{Germany}}
\email{butz@ifi.lmu.de}

\renewcommand{\shortauthors}{Amin et al.}

\begin{abstract}

As LLM conversations grow, their histories capture alternative directions, decisions, and evolving lines of thought that can be difficult to navigate through chat alone. We investigate an interaction concept that represents the same conversation through two synchronized views: a familiar linear chat for ongoing dialogue and a spatial canvas for navigating its emerging structure. To investigate this interaction concept, we developed CanvasConvo, which allows conversations to branch into alternative paths that remain accessible across both views. In a five-day field deployment with 24 participants, we examined how people appropriated this parallel representation in self-directed knowledge work. Participants selectively moved between the two views rather than replacing chat with the canvas. Chat remained central to conversational interaction, while the canvas supported overview, revisitation, and exploration of alternatives. Adoption was uneven, revealing challenges around established chat habits, transitions between representations, and understanding branch context. Our findings inform the design of user interfaces for LLMs that combine linear and non-linear conversation representations.
\end{abstract}




% \begin{teaserfigure}
%   \includegraphics[width=\textwidth]{figures/Screenshot 2026-03-30 at 12.54.49.png}
%   \caption{will change this, have my id in there :3. Planning on showing the process. AB: then make sure the process diagram is wide and not high, thus saving space ;-)}
%   \Description{...}
%   \label{fig:teaser}
% \end{teaserfigure}

\begin{teaserfigure}
  \includegraphics[width=\textwidth]{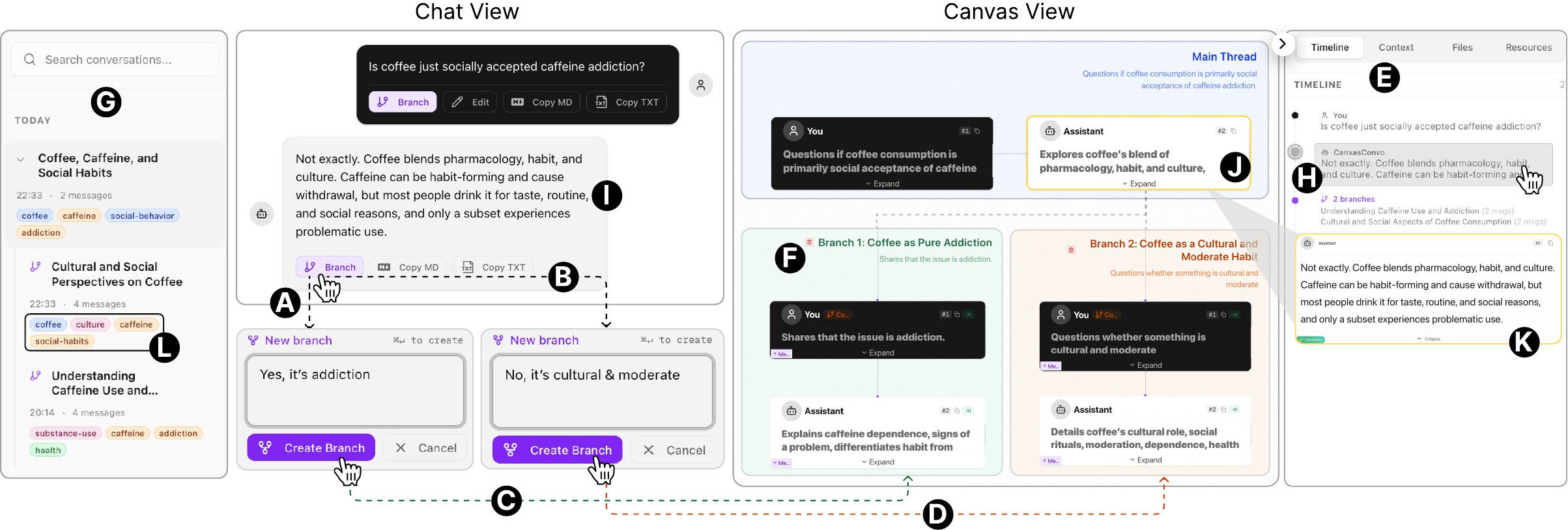}
  \caption{Our system CanvasConvo extends traditional chat-based interaction (Chat View) with branching and spatial organization (Canvas View). Starting from a single prompt, users can create alternative conversational paths directly from messages
  (A, B), enabling parallel exploration of different perspectives. The canvas view represents these branches as a structured node-link graph, supporting comparison, navigation, and revisiting of ideas (C, D). The timeline (E) is linked to both linear and spatial views: selecting an entry (H) scrolls to the corresponding point in the chat (I), while in the canvas, it highlights the referenced node (J) and enables users to zoom into its details on click (K). The sidebar (G) organizes conversations with tags (L).
  }
  % \caption{will change this, have my id in there :3. Planning on showing the process. AB: then make sure the process diagram is wide and not high, thus saving space ;-)}
  \Description{...}
  \label{fig:teaser}
\end{teaserfigure}

\received{20 February 2007}
\received[revised]{12 March 2009}
\received[accepted]{5 June 2009}

\maketitle

\section{Introduction}

Large language models (LLMs) have become integral to knowledge work, supporting activities such as writing, programming, planning, design, and research through iterative interaction~\cite{lee2024design}. During these tasks, users' goals and interpretations often emerge progressively through dialogue with AI rather than being fully specified upfront, making LLM interaction a process of intentmaking and sensemaking~\cite{intentmaking}. As these interactions unfold, conversations accumulate alternatives, constraints, and reusable strategies, becoming external representations of users' evolving thought processes~\cite{wordcraft,promptcanvas}. Yet LLM interaction remains dominated by linear conversational interfaces that become difficult to manage as conversations grow longer and more exploratory. Earlier ideas and context become buried, while exploring alternatives often requires copying prompts, rerunning conversations, or distributing work across multiple chats or tabs~\cite{Jiang_Rayan_Dow_Xia_2023,suhSensecapeEnablingMultilevel2023,abscribe,massonDirectGPTDirectManipulation2024}. As a result, users must increasingly manage not only the task itself, but also the accumulated context, alternatives, and progress of the interaction~\cite{metacognition}.

Researchers have explored richer interaction paradigms for addressing these limitations, including graphical representations of generated responses~\cite{Jiang_Rayan_Dow_Xia_2023}, executable prompting graphs~\cite{chainforge,prompt-chainer}, integrated rewriting environments~\cite{abscribe}, spatial canvases for organizing generated information~\cite{suhSensecapeEnablingMultilevel2023,suh2024luminate,promptcanvas}, and conversation-level branching~\cite{branchat}. These systems demonstrate the value of introducing structure into LLM interaction, but often require users to explicitly construct or manage that structure by assembling prompt graphs, arranging workspace artifacts, or organizing conversational branches. This leaves open how conversational structure can emerge directly from ongoing interaction while remaining integrated with the familiar chat experience. To ground this problem in users' existing practices, we conducted a formative study with eight experienced LLM users. The study identified recurring challenges around context fragmentation, iteration friction, and exploring alternative conversational paths, motivating the design of CanvasConvo.

Guided by these findings, we developed CanvasConvo, a conversational interface that combines conventional chat with a synchronized spatial representation automatically derived from the evolving conversation. Every message becomes part of a persistent conversational structure that users can revisit, branch, and organize without separately authoring a workspace or abandoning the familiar interaction model of chat. Rather than positioning the spatial canvas as a replacement for chat, CanvasConvo explores how linear and non-linear representations can coexist as complementary views of the same evolving conversation. Because the system is intended to support conversations that accumulate, branch, and are revisited over time, we sought to understand how these representations become incorporated into users' ongoing practices rather than evaluating only their initial use. Prior work shows that the qualities shaping initial encounters with interactive systems can differ from those that become important through continued use~\cite{karapanos}, while longitudinal studies of generative AI have identified processes such as familiarization, customization, and workflow appropriation that emerge through repeated interaction~\cite{userstudy1}. We therefore conducted a five-day exploratory field study with 24 participants who incorporated CanvasConvo into self-directed writing, programming, planning, research, and study activities. Using interaction logs, questionnaires, and semi-structured interviews, we examined how participants adopted and appropriated its conversational representations over time.

Our findings reveal a division of labor between the two synchronized representations: participants primarily treated chat as a generation surface to advance the conversation, while using the spatial canvas as an orientation surface to revisit, compare, and organize evolving conversational trajectories. Branching supported exploration of alternatives while preserving their relationship to the prior context, but also introduced new challenges in understanding which context a branch retained and in reconciling divergent conversational paths. Together, these findings suggest that supporting non-linear LLM interaction is not simply a matter of replacing linear histories with richer visual structures. Instead, conversational interfaces can benefit from treating linear and spatial representations as complementary interaction surfaces, supporting the fluid presentation of conversational content while making the accumulating structure of that interaction available for navigation, reflection, and exploration over time.

\section{Related Work}
% \done{redid sort of the whole RW}
We review prior work from three complementary perspectives: research on intent formation during conversational interaction with LLMs, approaches that structure LLM interactions beyond chronological chat, and spatial workspaces for supporting non-linear exploration. Together, these strands motivate our investigation of conversation-level spatial representations.

\subsection{Intent Formation in Conversational LLM Interaction}

Prior work characterizes interaction with LLMs as an iterative process in which users progressively formulate, evaluate, and refine their intent. Why Johnny Can't Prompt shows that non-expert users often develop prompts opportunistically, struggle to articulate their goals precisely, and lack systematic strategies for evaluating model outputs~\cite{johny}. Similarly, research on intentmaking argues that goals and interpretations emerge through interaction rather than being fully specified in advance~\cite {intentmaking}.

Several systems support this iterative process by externalizing intermediate conversational artifacts. Wordcraft enables users to iteratively rewrite, continue, and discuss generated text during collaborative writing~\cite{wordcraft}. PromptCanvas represents prompts and intermediate outputs as persistent, composable widgets that users can refine throughout creative workflows~\cite{promptcanvas}. Likewise, Selenite structures generated alternatives and evaluation criteria to support reasoning during exploratory decision making~\cite{selentine}. Although these systems address different domains, they collectively show that intermediate prompts, outputs, and decisions become valuable artifacts that support the gradual refinement of user intent.

\subsection{Structuring LLM Interaction}

Beyond supporting iterative intent formation, researchers have explored alternative representations for organizing interactions with LLMs. Existing systems structure different aspects of the interaction process, including generated responses, prompting workflows, conversational memory, and information spaces.

Some systems focus on structuring model outputs. Graphologue transforms generated responses into node-link graphs of entities and relationships, enabling non-linear exploration of generated content~\cite{Jiang_Rayan_Dow_Xia_2023}. VISAR represents argument development as an editable hierarchical tree, allowing users to organize and revise reasoning across multiple levels of abstraction~\cite{Zhang_2023}. Other work instead structures prompting workflows. PromptChainer and AI Chains decompose complex tasks into modular prompt pipelines with intermediate outputs that improve transparency and debugging~\cite{prompt-chainer,AI-chains}, while ChainForge represents prompt workflows as executable graphs for systematic comparison and evaluation~\cite{chainforge}. DirectGPT maps direct-manipulation operations to reusable prompts~\cite{massonDirectGPTDirectManipulation2024}, and Memolet externalizes conversational memory into reusable interactive objects~\cite{memolet}.

A complementary line of work structures broader information spaces assembled during exploration. Sensecape and Luminate organize generated information within hierarchical workspaces that support multilevel exploration and synthesis instead of committing users to a single conversational trajectory~\cite{suhSensecapeEnablingMultilevel2023,suh2024luminate}. Collectively, these systems demonstrate that explicit representations improve navigation, comparison, and reuse during complex interaction with LLMs, although they organize different artifacts of interaction rather than the conversation itself.

% \subsection{Spatial Workspaces for Exploratory Interaction}
\subsection{Spatial Workspaces and Non-Linear Interaction}

Spatial workspaces have become an increasingly common interaction paradigm for supporting complex cognitive tasks. Earlier research on conversation visualization showed that asynchronous discussions naturally evolve into multiple interleaved threads and benefit from persistent spatial representations that preserve context while enabling flexible navigation~\cite{convspace,forsense}. More recent work extends these ideas to interaction with LLMs.

PromptCanvas enables users to arrange prompts and generated outputs within a persistent workspace that supports iterative refinement through direct manipulation~\cite{promptcanvas}. Sensemaking systems such as CoNotate, InterWeave, and ForSense integrate adaptive guidance into user-created workspaces to support synthesis across evolving information structures~\cite{CoNotate,interweave,forsense}. DataTone similarly demonstrates how interactions within a workspace can be transformed into persistent constraints that shape subsequent analysis~\cite{datatone}. Recent systems have also explored branching conversational interaction. Branchat explicitly preserves alternative conversational trajectories, allowing users to revisit and continue previous lines of exploration without overwriting earlier work~\cite{branchat}. Together, these systems demonstrate the value of spatial organization and non-linear navigation for exploratory interaction with LLMs.

\subsection{Synthesis and Positioning}

Prior systems move beyond linear chat by structuring different parts of LLM interaction in different ways. As summarized in Table~\ref{tab:system_comparison}, these representations range from generated response content and prompt workflows to information spaces and conversation histories. They also differ in how structure is created and how it relates to chat: some require users to construct workflows or organize workspace artifacts, while others derive structure from the interaction itself; similarly, structured representations may replace, augment, or remain synchronized with conversational interaction. Together, these systems demonstrate multiple ways of externalizing structure to support exploration, navigation, and revisitation.

CanvasConvo focuses specifically on the conversation history as the object of representation. Its spatial structure is derived from the evolving conversation, with users creating alternative trajectories through branching, and remains synchronized with the familiar linear chat interface. Rather than replacing chat with a separate visual workspace, CanvasConvo treats linear and spatial representations as complementary views of the same conversation. While prior work has established the value of structured representations, comparatively less is known about how users incorporate such conversation-level representations into exploratory and long-running LLM workflows in practice.

Building on this gap, we examine both how users currently manage conversational complexity in existing LLM interfaces and how they subsequently interact with CanvasConvo during self-directed work. We focus on how users manage alternative directions, revisit and reuse prior conversational artifacts, and navigate non-linear structures during longer-running and multi-step tasks. We address the following research questions:

\begin{enumerate}[label=\textbf{RQ\arabic*}]
    \item How do users currently manage multiple ideas, alternative directions, or \textit{what-if} scenarios when interacting with LLM chat interfaces?
    \item How does providing non-linear conversational structures (e.g., branching, visual navigation, and reuse of conversational artifacts) influence users' ability to revisit, reuse, and adapt prior prompts and responses?
    \item How do users interact with such non-linear conversational interfaces when working on longer-running or multi-step tasks with LLMs?
\end{enumerate}

\begin{table*}[t]
\centering
\caption{Comparison of representative LLM interaction systems by primary
representation, structure creation, use case, relationship to chat, and
empirical evaluation.}
\label{tab:system_comparison}

\footnotesize
\setlength{\tabcolsep}{3pt}
\renewcommand{\arraystretch}{1.15}

\begin{tabularx}{\textwidth}{@{}
    l
    >{\raggedright\arraybackslash}X
    >{\raggedright\arraybackslash}X
    >{\raggedright\arraybackslash}X
    >{\raggedright\arraybackslash}X
    >{\raggedright\arraybackslash}X
@{}}

\toprule
\textbf{System} &
\textbf{Primary Representation} &
\textbf{Structure Creation} &
\textbf{Primary Use Case} &
\textbf{Relationship to Chat} &
\textbf{Evaluation} \\
\midrule

Graphologue~\cite{Jiang_Rayan_Dow_Xia_2023} &
LLM response content &
Automatic &
Information exploration &
Synchronized diagram and text &
Qualitative study \\

ChainForge~\cite{chainforge} &
Prompt workflow &
User-authored &
Prompt engineering / testing &
Separate visual workflow &
Lab study\\

PromptCanvas~\cite{promptcanvas} &
Prompt logic &
Hybrid &
Creative writing &
Canvas workspace &
2 lab studies + field study \\

Sensecape~\cite{suhSensecapeEnablingMultilevel2023} &
Information space &
Hybrid &
Exploration / sensemaking &
Structured workspace &
Lab study \\

Branchat~\cite{branchat} &
Conversation history &
Automatic + user branching &
Revisitation / context management &
Synchronized tree and chat &
Lab study \\

\midrule

\textbf{CanvasConvo} &
\textbf{Conversation history} &
\textbf{Automatic + user branching} &
\textbf{Exploratory and long-running LLM workflows} &
\textbf{Synchronized complementary canvas and chat views} &
\textbf{5-day field study} \\

\bottomrule
\end{tabularx}
\end{table*}

\section{Formative Study}

Prior work has identified challenges in managing increasingly long and complex interactions with LLMs, including difficulties in revisiting earlier content, exploring alternative directions, and maintaining context across conversational turns. However, less is understood about how users encounter and work around these challenges in their existing LLM workflows, particularly when conversations involve iterative exploration rather than a single linear progression.

To better understand these practices and inform the design of CanvasConvo, we conducted a formative study examining how users manage complex and exploratory interactions with current conversational LLM interfaces. The study focused on participants' existing practices: how they explore alternative directions, iterate on previous outputs, maintain context, and organize conversational histories as their interactions grow.

Eight expert LLM users (F1-F8), all with prior experience using conversational AI tools, were recruited through purposive sampling under the inclusion criterion of intensive professional LLM use (10+ sessions/week, at least 6 months of regular use); three were UI/UX researchers and five were software developers. Participants took part through in-person sessions, remote sessions, or asynchronous written feedback, depending on their availability covering workflow walkthroughs, breakdown moments, ideal-tool sketches, and reactions to early design directions. Working through the sessions together, the two researchers grouped the recurring observations into three core themes. 

% and informally checked their interpretations back with several participants. 

% The study is exploratory: it characterizes recurring friction in expert professional practice rather than estimating its prevalence or making statistical generalizations.

\subsection{Findings}
Participants expressed consistent frustration with the linear, sequential nature of current LLM interfaces. Despite their expertise, many lacked the vocabulary to articulate what they wanted until the session prompted structured reflection. \textbf{Three themes were identified as central issues. Context fragmentation, iteration friction, and the forking problem.} These themes align with research on information foraging \cite{pirolli2005sensemaking} and the tension between linear interaction paradigms and the non-linear nature of knowledge work \cite{suhSensecapeEnablingMultilevel2023}.

\subsubsection{Context Fragmentation} Context fragmentation refers to the scattering of relevant information across the conversation history, forcing users to scroll, search, or mentally reconstruct the state of their work. Participants consistently reported difficulty locating files and outputs, with F3 noting, \textit{``I spend half my time scrolling up to find the file I just uploaded. By the time I find it, I have lost my train of thought.''} Valuable outputs such as well-structured code snippets or polished paragraphs became embedded in the scroll history and were effectively inaccessible for later retrieval. F6 illustrated the resulting inefficiency: \textit{``I got this perfect function like 20 messages ago, and now I cannot find it. I have to scroll and scroll, and sometimes I just regenerate it, which is wasteful.''} Several participants also described the cognitive friction of session interruption. Upon returning after a break, they reported needing to rebuild their mental model of the entire conversation: \textit{``If I close the tab and come back an hour later, I have to re-read the whole thing to remember where I was''} (F1). This behavior reflects the extraneous cognitive load imposed by interfaces that provide neither persistent summaries nor spatial overviews of session state \cite{sweller}.

Six of eight participants had developed explicit workarounds: maintaining separate documents (Notion, Word, Google Docs) to preserve important outputs from the chat. This coping strategy echoes the principles of casual information organization, in which users create ad hoc collections of resources when the primary interface fails to support the necessary persistence of artifacts \cite{pile}. The widespread adoption of this workaround reveals a clear gap: critical information that should be persistent and retrievable must instead be manually externalized by the user. The design implication is a persistent, always-visible artifact repository that elevates uploaded files and generated outputs from embedded ephemera to first-class, directly accessible objects.

\subsubsection{Iteration Friction} 
Iteration friction refers to the disproportionate cost of making small, localized refinements to generated content \cite{10.1145/571985.571996}. Six of the eight participants described frustration with the binary choice between accepting an output as-is or regenerating it entirely. F2 articulated this constraint clearly: \textit{``I just want to tweak this one paragraph, not regenerate the whole essay. But if I ask it to fix paragraph three, it gives me a completely different version of the whole thing.''} To circumvent this limitation, users developed complex workarounds. F5 described constructing increasingly unwieldy prompts designed to protect desired portions while modifying others: \textit{``I end up writing these enormous prompts like `keep paragraphs 1, 2, and 4 exactly the same, but rewrite paragraph 3 to be more formal.' It is exhausting.''} More concerning was the inhibitory effect on iteration itself. F7 reported a reluctance to ask for revisions even when dissatisfied: \textit{``Sometimes I get something 90\% right, and I am scared to ask for changes because it might ruin what is already working.''} This fear-driven acceptance of suboptimal outputs represents a fundamental usability failure.

Seven of the eight participants reported manually editing LLM outputs rather than requesting revisions, and F3 described an extreme adaptation: copying individual paragraphs into entirely new chats for focused iteration, then manually recombining the results across sessions. These strategies highlight a critical design need: inline tools for localized iteration that allow users to target specific content segments for refinement without triggering full-context re-prompting or risking unintended side effects elsewhere in the output. This aligns with the research calling for a shift from purely conversational interaction toward direct manipulation of specific generated objects~\cite{massonDirectGPTDirectManipulation2024}. F2 raised an important concern about transparency: \textit{``How do I know the widget is only changing the selected text? What if it's also silently modifying the context around it?''} This does not invalidate the case for local iteration tools; rather, it identifies a requirement that must accompany them (preview affordances such as diff visualization or undo-group mechanisms) to calibrate user trust \cite{amershi}.

\subsubsection{The Forking Problem} 
The forking problem refers to the difficulty of exploring alternative approaches without losing accumulated context or polluting the conversation history. Seven of the eight participants encountered this tension between the desire to try new strategies and the cost of restarting from scratch. F8 quantified the friction: \textit{``If I want to try a different prompt strategy, I have to copy-paste everything into a new chat. That is like a five-minute tax every time I want to explore an alternative.''} Some users attempted to work around this by continuing in the same chat, but this strategy traded technical convenience for cognitive overhead. F6 described the result: \textit{``My chat looks like a graveyard of half-finished ideas. I cannot tell which approach I am currently on.''} Other participants adopted the multi-window strategy, opening parallel sessions to preserve context. However, this approach introduced new friction in the comparison phase. F1 reported: \textit{``I have to open two browser tabs side-by-side and manually compare. There is no way to see the branches together.''} These ad-hoc workarounds (mental models of scattered branches, copy-paste overhead, manual cross-tab comparison) point to a fundamental absence: integrated support for exploring and comparing parallel lines of inquiry within a single coherent interface, a capability essential for sensemaking with LLMs~\cite{Jiang_Rayan_Dow_Xia_2023, suhSensecapeEnablingMultilevel2023}.

The severity of this problem manifested in different ways across the sample. F4 reported habitually maintaining \textit{``15 to 20 ChatGPT tabs open at any time,''} each representing a distinct exploration thread. F2 adopted a more self-aware coping mechanism, typing explicit \textit{``BRANCH:''} markers directly into the chat to impose conceptual structure: a fragile strategy that offloads to working memory a distinction that the interface should maintain automatically.

These challenges point to a spatial canvas with integrated branching support. The cognitive benefits of stable spatial layouts for information retrieval and navigation are well-documented~\cite{10.1145/288392.288596}, and recent work shows that spatial organization can substantially reduce the overhead of managing complex, multi-threaded explorations~\cite{10.1145/2470654.2466430}. The design solution is to make branches visually salient, spatially distinct, and directly comparable.

This enthusiasm for spatial organization was not universal. F7 articulated a limitation: \textit{``I already have spatial organization in my editor. If I open yet another spatial interface for the LLM, am I coordinating two spatial models in my head, or does one replace the other? That feels complicated.''} Spatial interfaces may thus introduce friction when layered on existing spatially complex tools such as IDEs; their value is contingent on task complexity, workflow integration, and the user's broader tool ecosystem, a boundary condition the prototype accommodates through an optional embedded panel mode.

\subsection{Design Goals}

Our formative study revealed that users were already constructing implicit conversational branches through multiple chats, browser tabs, copy-pasting, and self-authored markers, creating a trade-off between preserving conversational context and keeping alternative trajectories manageable. Together with recurring challenges in retrieving and iterating on prior content, these observations, alongside prior work on non-linear LLM interaction, direct manipulation, and information organization, informed four design goals for supporting exploratory and long-running conversational workflows.

\begin{enumerate}[leftmargin=*, itemsep=1mm, label=\textbf{DG\arabic*}]

    \item\textbf{Support parallel exploration of conversational paths.}  
    % Interactions with LLMs often involve exploring alternatives, such as different ideas, approaches, or interpretations. 
    Prior work shows that users iteratively refine prompts and pursue multiple directions during creative and analytical tasks~\cite{johny, g-copilot}. However, conventional chat interfaces constrain interaction to a single linear trajectory. Grounded in the \textit{Forking Problem} identified in our formative study, the system should allow users to pursue alternative directions from existing conversational states without duplicating context or fragmenting their work across separate chats. These alternatives should remain connected to their shared conversational history so that users can move between and compare different lines of exploration.
    
    % The system should enable users to create and explore multiple conversational paths in parallel without losing prior context.

%AB I would say DG2 is just a specific solution to DG1 and DG3 and thus redundant.
%    \item [\textbf{DG2}] \textbf{Externalize conversation structure as a spatial and persistent representation.} Conversational interfaces typically embed structure implicitly within a scrollable history. \textcolor{blue}{Research in sensemaking and external cognition highlights the importance of making intermediate states visible and manipulable~\cite{}}.     The system should represent conversations as explicit structures, such as a branching conversation tree within a spatial canvas, allowing users to inspect, navigate, and organize interactions over time.

    \item \textbf{Enable direct manipulation of conversational artifacts.}  
    Informed by the \textit{Iteration Friction} observed in our formative study, the system should enable users to act on specific conversational elements rather than relying exclusively on sequential prompting and full-response regeneration. Following principles of direct manipulation, conversational artifacts should remain persistent and operable~\cite{massonDirectGPTDirectManipulation2024}, allowing users to selectively reuse, branch from, compare, or transform prior content.

    \item \textbf{Support scalable navigation and context management.}  
    % As conversations grow in length and complexity, users face challenges in maintaining awareness of prior interactions and retrieving relevant information. 
    % % Prior work explores summarization, tagging, and visualization to support navigation~\cite{cheng2022binding, park2023generative}.  
    % The system should provide mechanisms for navigating conversational structures at multiple levels of detail, supporting both overview and focused interaction following Shneiderman's mantra~\cite{shneiderman1996eyes}.
    Grounded in the \textit{Context Fragmentation} identified in our formative study, the system should help users maintain awareness of conversation structure as interactions accumulate. Users should be able to move between an overview of their conversational history and focused interaction following Shneiderman's mantra~\cite{shneiderman1996eyes} with particular branches, messages, or artifacts, supporting both orientation and retrieval in increasingly complex conversations.

    \item \textbf{Balance structured guidance with flexible interaction.}  
    Throughout the formative study, participants developed different strategies for managing complex LLM interactions, including using external documents, multiple chat windows, and self-authored markers to organize branches. Structured interaction techniques, such as prompt suggestions and scaffolding, can support users in exploring and refining ideas~\cite{johny}. At the same time, overly rigid systems may constrain user agency. The system should provide context-aware guidance while preserving flexibility in how users explore, structure, and manipulate conversations.

\end{enumerate}

\section{CanvasConvo}
\begin{figure*}[t]
    \centering
    \includegraphics[width=\linewidth]{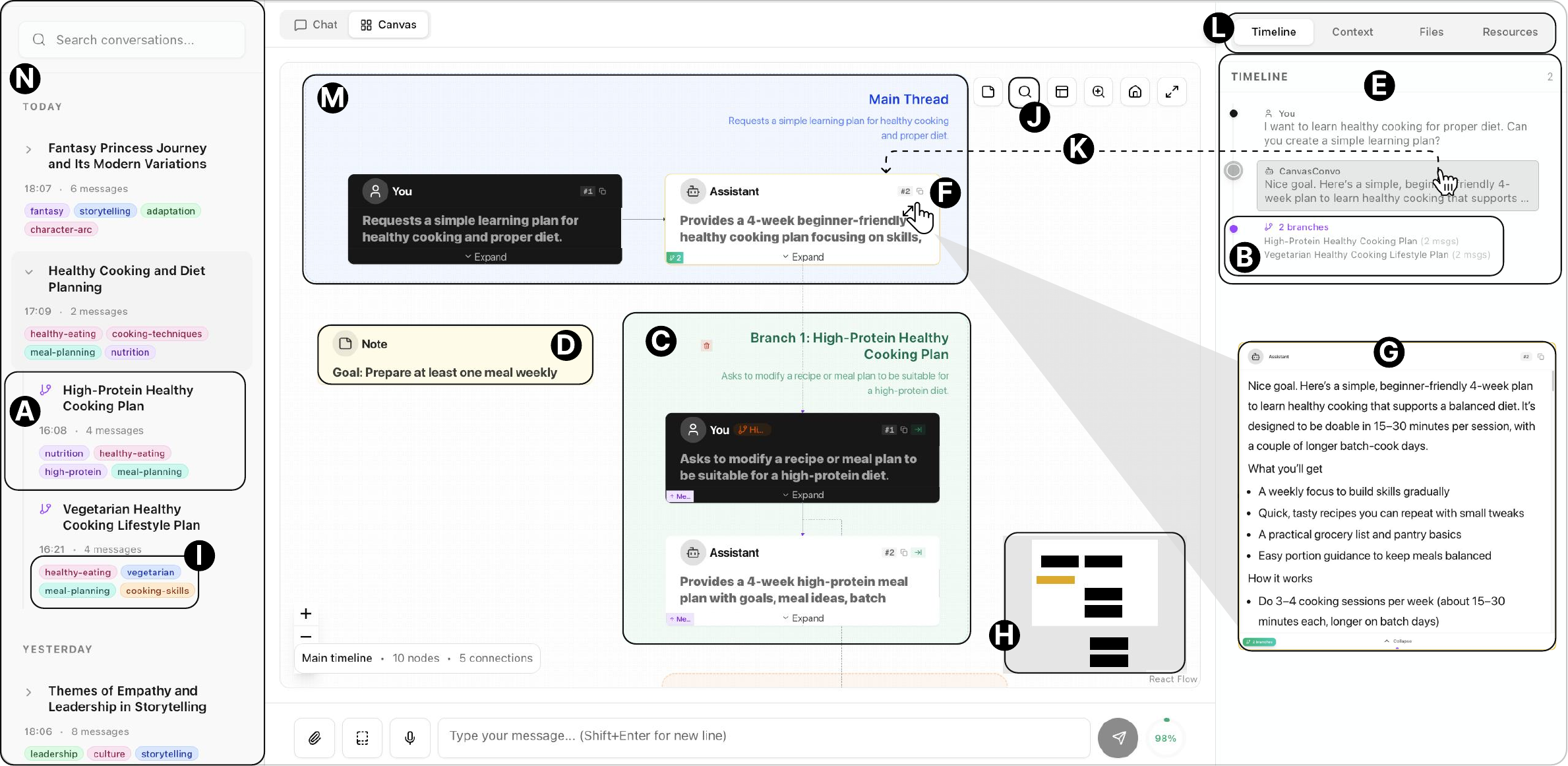}
    % \caption{Canvas View}
    \caption{CanvasConvo (canvas view). The canvas (center) visualizes messages as nodes with a linear main thread (M), and branching zones (C, F), while the sidebar (N) organizes conversations (with tags (I)) and branches (A). Users can annotate with notes (D), navigate via semantic zoom (G) and a minimap (H), and search (J) to highlight relevant content. The timeline (E) shows conversation history and enables navigation to messages (K), with branch points explicitly marked (B). The overview panel (L) displays contextual information, including goals and resources.}
    \Description{...}
    \label{fig:interaction1}
\end{figure*}

Building on our design goals, CanvasConvo conceptualizes interaction with LLMs as a \textit{non-linear workflow} rather than a sequence of isolated turns. 

\subsection{Interaction Concept: Structuring Non-Linear LLM Workflows}
% \done{Added the table}
% \done{The primary design choices make sense, but there are several additional features that do not feel grounded with the core problem the authors are trying to tackle.

% - For example: 3C - The toolbar’s feature makes sense for a writing task, but it is hard to understand what design goal it is related to.}

Inspired by interaction paradigms from visual editing systems and prior work on structured writing environments, CanvasConvo combines a familiar chat interface with spatial and structural interaction mechanisms. The system organizes interaction across three complementary dimensions: spatial organization, branching, and temporal navigation, allowing users to fluidly move between focused dialogue and exploratory workflows.

% \textcolor{blue}{\autoref{tab:design-mechanisms} maps each mechanism to the design goal it serves. We distinguish core mechanisms, which realize the concept, from supporting mechanisms, which address friction we observed in the pre-study but are not themselves claims of the paper.}

% \begin{table}[t]
% \centering
% \caption{\textcolor{blue}{CanvasConvo mechanisms categorized as core contributions or supporting functionality, and their relationship to the design goals.}}
% \label{tab:design-mechanisms}
% \begin{tabular}{p{7.3cm}cc}
% \toprule
% \textbf{Mechanism} & \textbf{Type} & \textbf{Goal} \\
% \midrule
% Branching from any message                  & Core       & DG1, DG5 \\
% Derived canvas / node-link view            & Core       & DG1, DG3 \\
% Bidirectional chat $\leftrightarrow$ canvas sync & Core & DG1, DG3 \\
% Timeline with marked branch points         & Core       & DG3, DG5 \\
% \midrule
% Semantic zoom, minimap, search             & Supporting & DG3 \\
% Sticky notes on canvas                     & Supporting & DG2 \\
% Tags and automatic summaries               & Supporting & DG3 \\
% Prompt templates                           & Supporting & DG2, DG4 \\
% Context \& resource panel                  & Supporting & DG5 \\
% Selection toolbar (shorten, rephrase, \ldots) & Supporting & DG2 \\
% \bottomrule
% \end{tabular}
% \end{table}

\subsubsection{Spatial Canvas}

The spatial canvas serves as a structural representation of the conversation, transforming the linear message sequence into a node-link graph. Each message is represented as a node, and edges encode sequential relationships and branch connections. The layout is \textbf{system-derived and structure-driven}. The main conversation thread is arranged linearly (Main Thread, \autoref{fig:interaction1}-\blackcircled{M}), while branches appear as separate zones connected to their origin points. The layout reflects the underlying conversational structure. The canvas introduces several key interactions: 

\begin{itemize}[leftmargin =*]
    \item \textit{Node Types.} Messages are represented as chat nodes containing role labels and content previews (\autoref{fig:interaction1}-\blackcircled{F}). Branches are grouped into \textit{branch zone nodes}, visually distinct regions with labels and automatically generated summaries (\autoref{fig:interaction1}-\blackcircled{C}). Additional node types include loading placeholders and user-created sticky-notes for annotations (\autoref{fig:interaction1}-\blackcircled{D}). With these, users can add annotations, for example, to externalize ideas, without modifying the conversation.

    \item \textit{Semantic Zoom.} The canvas supports four levels of semantic zoom (Cluster, Minimal, Compact, Full), from high-level role indicators to full message content (\autoref{fig:interaction1}-\blackcircled{G}). This enables users to move between overview and detailed inspection.

    \item \textit{Navigation and Search.} Users navigate via panning, zooming, keyboard shortcuts, and a minimap (\autoref{fig:interaction1}-\blackcircled{H}) that provides a global overview. A text search function (\autoref{fig:interaction1}-\blackcircled{J}) highlights relevant nodes by adjusting opacity, allowing quick identification of content in large conversations. Hovering over any node in the canvas further highlights the corresponding conversational path.

    % \item \textit{Annotations.} Users can add sticky-note annotations to the canvas, enabling lightweight externalization of ideas without modifying the underlying conversation.
\end{itemize}

Overall, the canvas functions as a \textit{visualization and navigation layer} that makes conversational structure, relationships, and alternative paths explicitly visible \textbf{(DG3)}.

\subsubsection{Conversation Branching}

CanvasConvo introduces branching as a core mechanism for exploring alternative conversational trajectories. From any message, users can create a branch that duplicates the conversation history up to that point and continues as a new thread (\autoref{fig:interaction2}-\blackcircled{G}). This preserves prior context while enabling divergence. Branch creation is integrated directly into the interaction flow via buttons in chat messages and canvas nodes. Branches are represented consistently across the views:
\begin{itemize}[leftmargin =*]
    \item as nested entries in the conversation sidebar (\autoref{fig:interaction1}-\blackcircled{A})
    \item as badges attached to branching points in the chat (\autoref{fig:interaction2}-\blackcircled{A})
    \item and as distinct zones in the spatial canvas (\autoref{fig:interaction1}-\blackcircled{C})
    % \item highlighted in different color in the timeline (\autoref{fig:interaction1}-B)
\end{itemize}

Hovering over a branch tag reveals a preview of the branch (\autoref{fig:interaction2}-\blackcircled{B}). Branching supports workflows such as exploring alternative prompts, comparing different outputs, or recovering from earlier points in the conversation. By externalizing alternative paths, branching transforms interaction from a single linear progression into a \textit{multi-path exploration process} \textbf{(DG1)}. It is also possible to delete branches if needed.

\subsubsection{History Timeline}

\begin{figure*}[t]
    \centering
    \includegraphics[width=\linewidth]{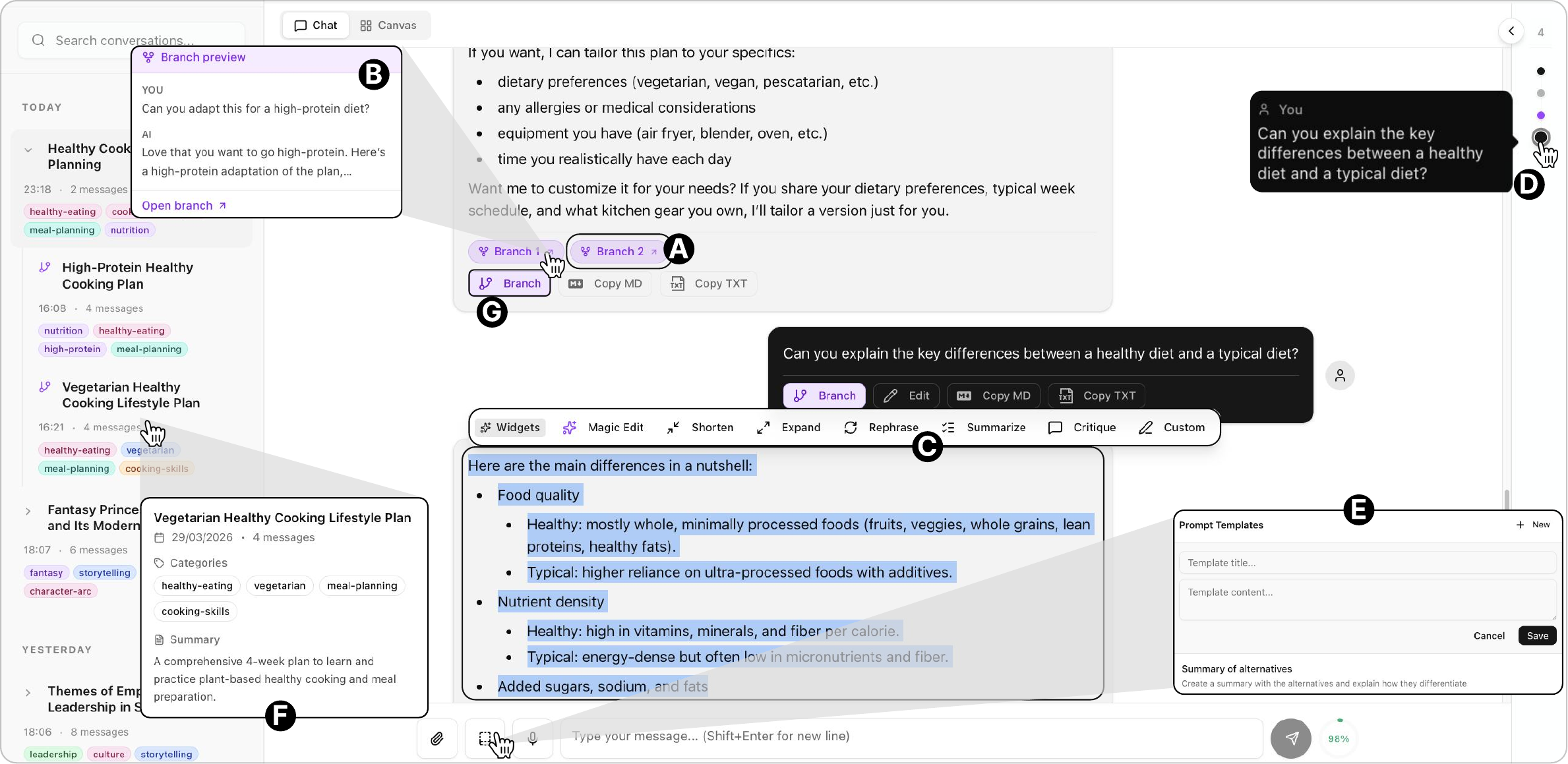}
    \caption{CanvasConvo (Chat view). Messages can be branched directly (A), with previews shown on hover (B) and branch points indicated in the thread (G). Hovering over a point in the timeline shows an overview of the conversation (D). A contextual toolbar (C) enables inline transformations such as summarizing or rephrasing selected text. The sidebar provides conversation overviews and metadata (F), while prompt templates (E) support reusable inputs.}
    \Description{...}
    \label{fig:interaction2}
\end{figure*}

To complement the structural view provided by the canvas, CanvasConvo includes a timeline view within the overview panel. It presents a vertical history of all messages, with each entry represented by a colored marker and a text preview (\autoref{fig:interaction1}-\blackcircled{E}). Selecting an entry navigates the viewport to the corresponding message in both chat and canvas view (\autoref{fig:interaction1}-\blackcircled{K}). Branch-point messages are explicitly indicated (\autoref{fig:interaction1}-\blackcircled{B}). Hovering over a timeline entry reveals a preview of the corresponding conversation text (\autoref{fig:interaction2}-\blackcircled{D}, there shown in the collapsed state of the timeline sidebar). This view emphasizes \textbf{temporal progression \textbf{(DG3)}}, allowing users to:
\begin{itemize}[leftmargin =*]
    \item revisit earlier states of the conversation
    \item track how ideas evolved over time
    \item and quickly navigate across different points in the interaction
\end{itemize}

Together with the canvas, the timeline provides a complementary perspective: while the canvas shows \textit{structure and relationships}, the timeline reveals \textit{sequence and history} by preserving the chronological flow of both user and assistant messages.

\subsubsection{Supporting Interaction Mechanisms}

% \begin{figure}[t]
%     \centering
%     \includegraphics[width=0.6\linewidth]{figures/context.pdf}
%     \caption{Chat View}
%     \Description{...}
%     \label{fig:context}
% \end{figure}

In addition to these core components, CanvasConvo provides several mechanisms that support interaction across views and with chat messages in general:

\begin{itemize}[leftmargin =*]
    \item \textit{Chat Interface.} This view shows messages with real-time streaming responses and rich rendering (Markdown, code, LaTeX, images). Messages support actions such as edit, regenerate, copy, and branch. The sidebar on the left ((\autoref{fig:interaction1}-\blackcircled{N})) shows a searchable list of conversations by recency; each displays metadata such as message count and conversation tags (\autoref{fig:interaction1}-\blackcircled{I}). The tags are generated based on the whole context of the conversation and dynamically updates as the conversation progresses. Hovering reveals a summary preview (\autoref{fig:interaction2}-\blackcircled{F}). At the bottom, the message input supports reusable prompts via a picker \textbf{(DG4)}, to quickly insert and reuse templates (\autoref{fig:interaction2}-\blackcircled{E}). %After each assistant response, the system generates context-aware suggestions (e.g., deepen, compare, simplify), for users to quickly continue in different directions.

    \item \textit{Toolbar.} Selecting text within a message reveals a %context-aware 
    toolbar with transformation actions (e.g., shorten, expand, rephrase, summarize, critique), see \autoref{fig:interaction2}-\blackcircled{C}. When clicking one, a popup shows a preview for users to confirm. %Transformations are previewed before application, 
    These tools enable localized refinement of outputs without writing new prompts \textbf{(DG2)}.
    % \rifat{accept or reject pop up eith the changes...}

    \item \textit{Context and Resource Management.} The overview panel provides access to contextual controls (e.g., inferred goals, system instructions), uploaded files, and automatically extracted resources such as links and generated images (\autoref{fig:interaction1}-\blackcircled{L}), collapsed in \autoref{fig:interaction1}. This makes otherwise implicit context visible and configurable.
\end{itemize}

\subsection{Technical Implementation}
CanvasConvo is implemented as a single-page web application using Next.js (App Router) and React, with Supabase providing authentication, persistence, and real-time synchronization. The spatial canvas is built on React Flow. Model responses are streamed via Server-Sent Events to enable low-latency interaction. The system integrates multiple LLMs %organized into tiers 
(e.g., GPT-5, GPT-4.1, GPT-4o), with configurable reasoning effort and large context windows supporting extended interactions.

\section{Main Study Design}
%AB shorten here...

% \done{authors were interested in studying how users managed workflows during “LLM-based writing activities” writing task. But section 5.2, 5.3 state that participants engaged in open-ended usage. This contradicts with the writing in Section 5 about the study’s goals.}

We conducted an exploratory study to investigate how users interact with our system. We were interested in understanding how users managed long-running LLM-supported workflows as they incorporated CanvasConvo into their everyday tasks. Participants were free to use the system for their own writing, programming, planning, research, study, and other knowledge-work activities, allowing us to observe how conversation-level spatial representations were appropriated across naturally occurring workflows. The study was approved by our institution's ethics board.

\subsection{Participants}
We recruited 24 participants (P1-P24, with no overlap with participants from the formative study) through institutional mailing lists and word-of-mouth (6 female, 17 male, 1 non-binary) with ages 21 to 54 years ($M = 28.6$, $SD = 6.6$). They represented diverse professional backgrounds, including students, researchers, software engineers, and industry professionals. Participants varied in their writing experience, including both professional and non-professional writers. All reported prior experience with LLM-based tools, with most indicating high familiarity and frequent use in their daily workflows. Participants commonly used LLM systems for tasks such as writing and editing, coding, data analysis, brainstorming, planning, and learning. Overall, the participant pool reflects a group of active LLM users with varied domains of application and experience levels. A detailed overview is provided in \autoref{tab:participant-overview2}. Participants were compensated 20€ for the five-day study.

\subsection{Study Procedure}
Our study allowed participants to integrate the system into everyday workflows. The procedure combined a structured onboarding task with open-ended usage.

\subsubsection{Pre-Task Preparation}
Before starting the study, participants completed an introductory questionnaire, watched a tutorial video\footnote{submitted as a supporting material to this submission} introducing the system, and got access to CanvasConvo. The questionnaire gathered data on participants' experiences with multi-step workflows in LLM chats, including how they structure complex tasks, manage multiple lines of thought, reuse prompts, and organize conversations. We further assessed perceived challenges, satisfaction with current interfaces, and desired features for supporting parallel exploration and long-running tasks. 

\subsubsection{Guided Mandatory Task}
Participants first completed a structured, mandatory task during the five-day study period. They selected or defined a topic such as a learning plan, weekly routine, creative writing idea, or personal project. They created an initial version of their idea through interaction with the AI, iteratively refined it via follow-up prompts, and explored alternative versions by modifying aspects such as difficulty, constraints, or style. Participants were encouraged to create multiple branches,  reuse prior prompts, revisit earlier responses, and use the canvas view to understand the conversation structures.

\subsubsection{Open-Ended Usage}
% After completing the onboarding task, participants used the system for their own tasks over the remaining study period. People used it for activities such as writing, planning, brainstorming, studying, or problem-solving.

After completing the guided task, participants used the system freely for their own work during the remainder of the five-day study. These activities included planning, programming, research, ideation, and other self-directed LLM-supported workflows, allowing us to observe adoption of the interface under naturalistic usage conditions.

% \done{added explanation: It would be nice to see what tasks participants actually completed during open-ended usage Understanding which task types drove the longest canvas sessions would help to frame the contribution better.}

\subsubsection{Post-Task Measures and Interview}
At the end of the study, participants completed a post-task questionnaire on usability, task load, and questions on workflow support and conversation management, and participated in a semi-structured interview. The interviews focused on participants' experiences, workflows, perceived benefits and challenges, and suggestions for improvement.

% At the end of the five-day study, participants completed an end-of-study questionnaire consisting of the System Usability Scale (SUS), the Raw NASA Task Load Index (NASA-TLX), and custom Likert-scale questions assessing workflow support, revisiting and reuse, management of multiple conversational directions, and sense of control and authorship. Participants also took part in a semi-structured interview focusing on their experiences, workflows, perceived benefits and challenges, and suggestions for improvement.

% \done{In Section 5.2.4, there is a mention of a post-task questionnaire. It is unclear what the post-task questionnaire was about.}

\subsection{Measurements and Analysis}

We collected both quantitative and qualitative data to capture user behavior and perceptions.

\subsubsection{Quantitative Measures.} Participants completed standard usability and workload instruments, including the System Usability Scale (SUS)~\cite{Brooke_96} and NASA Task Load Index (NASA-TLX)~\cite{hart1988development} along with custom Likert items on workflow support, revisiting and reuse, managing multiple directions, and sense of control and authorship. We also logged interaction data, including branching behavior, node creation, navigation patterns, and reuse of conversational artifacts.

\subsubsection{Qualitative Analysis.} 
% We analyzed open-ended survey responses and interview transcripts using inductive coding. Two authors performed open coding to identify themes around workflows, exploration, and interaction, and iteratively grouped and refined the codes with co-authors.
% \done{write that they were semi-structured like branchvis}
Interview recordings were transcribed and independently reviewed by two researchers. Because the interview protocol was organized around topics aligned with our research questions, including participants' existing conversational workflows, navigation and revisitation, exploration through branching, and experiences with the spatial representation, the analysis focused on identifying recurring patterns within and across these discussion areas. The researchers iteratively compared their observations, grouped related responses, and discussed emerging patterns and differences until reaching consensus. The resulting themes were used to organize the qualitative findings, with representative quotations used to illustrate both recurring experiences and divergent cases.

\section{Results}
%\rifat{objective telemetry}
%\rifat{add intro}
%We present findings from our main study of 24 participants. Qualitative data from post-study surveys and interviews examines how participants structured their workflows, used non-linear features, and reflected on their experience with the system. Objective interaction telemetry further captured feature engagement, branching behavior, and spatial navigation patterns across the study period. 
% We report the qualitative findings first, from existing experiences with usual tools to experiences with the new design, followed by the analysis of interaction telemetry.

We report the qualitative findings first, beginning with participants' existing
practices in conventional LLM interfaces and then examining how they adopted,
adapted, and encountered challenges with CanvasConvo during the five-day
deployment. We subsequently relate these statements to patterns observed in the
interaction telemetry.

% \daniel{We might get away with it for UIST specifically, but it would be stronger if we can report in more detail how we conducted the qualitative analysis. We could also strategically leave it out for now and respond to it in a rebuttal if it comes up -- but in any case, we should be able/prepared to articulate this process.}

% \done{Several quotes signal substantive problems with the design: P19 reports never using the canvas due to habit ("I just scrolled up... I never ended up using the canvas due to force of habits"), P20 found switching between canvas and chat confusing, and P23 reports a fundamental confusion about branch semantics ("I didn't know what the branch knows and what it doesn't know"). The telemetry also shows chat interactions dominating over canvas interactions (47.2\% vs. 28.3\%) and only 5 of 24 participants ever creating a prompt template:

% - engage with what these signals mean for the design rather than framing the study as broadly positive}

% - Branching semantics were unclear to multiple participants (""I didn't know what the branch knows and what it doesn't know"" P23, P20). I would expect to see either a redesign response or a discussion of how this problem should be solved.

\subsection{Practices and Challenges in Conversational UI Workflows (RQ1)}
\label{sec:prestudy-results}
Our findings from the introductory questionnaire (\textit{before} seeing/using our design, separate from the formative study) reveal how participants currently structure, manage, and struggle with workflows in their usual LLM chat interfaces. %We identify recurring patterns in how users organize tasks, navigate multiple ideas, and cope with limitations of existing systems.

\subsubsection{Structuring Multi-Step Tasks}

Participants described breaking complex tasks into smaller steps and iterating incrementally. They followed structured, step-by-step approaches, often starting with a high-level plan before refining individual components: \textit{``I usually ask one chat for a plan of the complex task… and then start new chats for each task.''}--P15. Others described iterative refinement within a single thread: \textit{``I outline the text… draft with AI… give feedback… iterate.''}--P1. Several participants emphasized decomposition and incremental problem solving: \textit{``I go step by step, with a bottom-up approach… solving a smaller problem first.''}--P5. However, not all workflows were structured. Some participants reported more organic and less controlled approaches: \textit{``I usually start chatting and work through the task organically… my workflow can sometimes feel unstructured''}--P17.

\subsubsection{Managing Conversations: One Thread vs. Multiple Chats}
Participants described their typical use of LLM chat interfaces as relying on a single, continuously growing conversation thread, which often became difficult to manage over time 
% (P2, P7, P17, P19)
. For example, P7 noted that they usually work within \textit{``one long thread.''} As conversations grew, this structure led to a loss of clarity and increasing complexity. P2 described how they would \textit{``fill it with so much unnecessary paths''} and eventually need to restart the chat altogether. Others characterized this experience as disorienting, with P14 noting that one can \textit{``go down a rabbit hole… and not really know how you got there.''}. To cope with these limitations, participants reported switching between multiple chats or restarting conversations entirely, leading to fragmented workflows and loss of context. Some split tasks across multiple chats to improve clarity: \textit{``I split them… to have a better overview of the `big task'.''}--P16. A common strategy combined both approaches. For example P10 noted, \textit{``If the topic can be split… I split them up in multiple chats''}. This highlights a tension between context continuity and organizational clarity, with no single strategy fully addressing both.

\subsubsection{Managing Multiple Ideas and Exploration}

When exploring alternative directions, participants relied on ad-hoc strategies such as scrolling, copying content, or opening multiple chats or browser tabs: \textit{``Continue in the same chat and scroll back… copy prompts into another chat''} --P11. Some used external tools to compensate for missing functionality: \textit{``I usually write external notes… to avoid unnecessary context.''}--P15. Despite these efforts, participants reported limited support for parallel exploration: \textit{``One thread can't always handle different tasks… I have to start another thread.''}--P7.

\subsubsection{Loss of Overview in Long Conversations}

A common challenge was maintaining an overview in longer interactions, due to the difficulty of retrieving earlier information or understanding past decisions. P8 said, \textit{``It is hard to find previous answers or go back and build upon them''}. Similarly, P10 said, \textit{``When the chat gets longer, it is difficult to keep track of what I discussed before''}. Participants attributed this to the linear, scroll-based nature of chat interfaces, such as P11:
\textit{``Scrolling a lot to previous chats.''} %This breakdown of overview becomes particularly problematic in multi-step and exploratory workflows.

\subsubsection{Need for Structure, Navigation, and Branching}

In their reflections on their usual tools, people consistently expressed the need for better organization and navigation, such as the ability to %branch conversations and 
explore alternatives without losing prior context. P12 mentioned, \textit{``I would add a way to create separate threads inside the same chat… to explore different ideas without mixing everything''}. P23 noted, \textit{``There is no true ‘undo' or branching feature… you're often forced to start over''}. Other desired features included grouping, visual overviews, and improved navigation: \textit{``A visual overview or map of the conversation''}--P8. P10 mentioned, \textit{``Organizing the chat into different sections or topics with an overview''}.

Participants also highlighted the need for better retrieval and reuse: \textit{``Being able to save certain prompts to use them afterwards.''}--P15.
Overall, these findings point to limitations of linear chat UIs in multi-step and exploratory workflows, and thus confirm that our design features address a user need.

\subsection{Effects of Non-Linear Structures on Reuse, Navigation, and Exploration (RQ2)}
\label{results:explorationbranching}
We next report the qualitative findings from participants' experiences with CanvasConvo, specifically on how the non-linear structures support revisiting, reuse, navigation, and exploration.

% \subsubsection{Transition from Linear Chats to Structured Workspaces}

\subsubsection{Adopting Spatial Interaction Alongside Linear Chat}

%Participants consistently contrasted CanvasConvo with traditional linear chat interfaces, describing a transition toward explicitly structured conversational workflows. 
Participants contrasted CanvasConvo with traditional chat interfaces.  Participants described preferring traditional chat interfaces for quick, low-effort interactions. For example, P3 noted that \textit{``for very minimal searches… I would use the other ones''}, while P1 stated \textit{``for quick things… I would just use a normal chat''}. Some continued to rely on linear interaction habits (P19, P23): \textit{``I just scrolled up… I never ended up using the canvas due to force of habits.''} (P19), others characterized their prior LLM use as sequential and difficult to organize in contrast, 
% (P17, P19, P20, P23). 
such as P17: \textit{``I normally use LLM tools in a very linear way, just in a long chat.''}

These statements indicate that adopting the spatial representation was not
immediate or uniform. For some participants, established chat-based practices
remained the default even when alternative navigation mechanisms were available.
In particular, P19's continued reliance on scrolling suggests that introducing a
spatial representation does not by itself displace interaction habits developed
through conventional chat interfaces. Participants instead differed in when and
for which kinds of tasks they considered the additional structure worthwhile.

In contrast, participants described CanvasConvo as supporting exploration and organization: 
% (P11, P17, P19)
P11 described it as \textit{``the go-to solution when I'm learning something,''} and P1 highlighted its usefulness \textit{``for longer ideations… trying out different topics… the branching is really nice.''} More broadly, participants pointed to its applicability in larger-scale tasks, as reflected by P23's observation that \textit{``people working on bigger projects would benefit the most.''} And P7 said, \textit{``usually… I have to use one long thread… but here… the organization was better and time efficient.''} They adopted spatial metaphors, framing the system as a map or structured overview: 
% (P11, P17, P19)
 % \textit{``I understand it as a map of the conversation.''} (P17).
 \textit{``like a mind map… where your ideas branch out.''} (P6). P3 perceived it as a tree, saying \textit{``it's kind of like a flow chart or like a tree''}.

% These findings suggest that the design features supported participants in non-linear exploration, beyond their usual LLM chat interactions, as contrasted by them.

Together, these findings suggest that participants did not simply replace linear
chat with a spatial workspace. Rather, they selectively incorporated spatial
interaction when conversational complexity made the additional structure useful,
while continuing to rely on familiar linear interaction for shorter or simpler
tasks.

\subsubsection{Externalizing Conversation State and Supporting Navigation}

Participants highlighted how the canvas reduces the need to rely on memory or scrolling, by visualizing branches and conversation states: \textit{``You can just click on older conversations… it was very easy to navigate.''}--P11. P17 noted, \textit{``I could clearly see where I had to go and how the conversation developed''}. Participants described this as enabling faster comprehension of prior work, such as P19: \textit{``It gives you a much faster way to get a feeling of what has happened''}. P9 explicitly compared the interaction to human conversational behavior: \textit{``It's more like how we talk to people… you can go back to something you mentioned earlier and continue from there''}. Participants described the canvas as making prior conversational states directly accessible rather than requiring them to reconstruct context from a linear history. Instead of rereading long transcripts, they used the timeline to jump to earlier moments in the conversation and the spatial layout to distinguish alternative reasoning paths. As P10 explained, ``There was a timeline on the right side... I could just hover over the timeline dots'' to revisit previous parts of the conversation, while DP6 highlighted the ability to move fluidly between ```the overview'' and detailed content. These accounts suggest that the canvas supported navigation not simply by making conversations easier to browse, but by externalizing conversation state into a persistent spatial representation that reduced the need to mentally reconstruct where ideas originated and how they evolved.

% Overall, these findings suggest that externalizing conversational structure on the canvas supported navigation. 

% \done{The section ends with the statement that the findings supported navigation, but the quotes don’t convey much about the process beyond saying it was “easy to navigate”.}

% These findings indicate that while externalization does improve awareness, effective retrieval mechanisms (e.g., search, labeling) remain critical.

\subsubsection{Branching as a Mechanism for Exploration and Control}

% The central mechanism for non-linear exploration was branching, although its semantics were not always clear (P23, P20): \textit{``I didn't know what the branch knows and what it doesn't know''} (P23). 

The central mechanism for non-linear exploration was branching, but participants
did not always develop an accurate mental model of its semantics. P23, for
example, explained: ``I didn't know what the branch knows and what it doesn't
know.'' This uncertainty concerned the conversational context inherited by a
branch rather than the mechanics of creating one. It suggests that making
alternative paths visually explicit is insufficient if the relationship between
their underlying conversational contexts remains ambiguous.

Participants used branching to explore alternatives, separate ideas, and avoid interference between different lines of reasoning. 
% (P2, P10, P17, P19, P11)
P19 said, \textit{``I used the branching tool to explore different ideas and evaluate them''}. P7 used branching in a highly structured, task-oriented way, saying \textit{``I made one branch for… materials… another branch for… logistics, guest list, and to-do things.''}. Participants also emphasized control and independence between branches 
% (P17, P19)
. P10 explained, \textit{``The branching kind of gave me an incentive to explore more directions.''} 
These findings show a tension in branching interactions: branches provided participants with explicit spaces for exploring alternatives, but their usefulness depended on participants' understanding of which conversational states were shared or isolated across those spaces.
% This suggests that branching supports exploration and control, and in doing so might also lower the threshold for trying alternative ideas in conversations with a chat interface.

% Participants further identified strong potential for CanvasConvo in complex workflows 
% % (P11, P17, P23)
% : \textit{``I would use it for planning, brainstorming, structuring ideas.''} (P17). \textit{``People working on bigger projects would benefit the most.''} (P23).

% However,  This highlights a key tension: while branching enables powerful exploration, it requires clear mental models and interface support to be effective.

\subsection{Interaction with Non-Linear Interfaces in Multi-Step and Long-Running Tasks (RQ3)}
\label{results:spatialStructure}
In the open-ended part of the study, participants applied CanvasConvo to a variety of tasks over several days, including planning, programming, research, and ideation. Insights from these experiences extend and substantiate the previous aspects with concrete instances across real tasks.

\subsubsection{Supporting Parallel and Iterative Workflows}

% \rifat{6.3.1,  participants resorted to using chat for the most part and the non-linear representation does show an overview of the conversations differently. But the section that is supposed to delineate how users manage parallel workflows:

% - Since this is a core aspect of the study, it is important to bring in evidence to help understand the frictions that emerge there.
% - Though the quotes conveyed what they did, it does not tell much about the processes behind managing parallel workflows.}

Participants described using branching to explore alternatives in parallel and structure their work across multiple directions. Rather than progressing sequentially, they distributed related ideas across branches and worked on them simultaneously. As P5 explained, \textit{``Well, I'm also looking at something else in parallel, so that way I speed up my work''}. Similarly, P6 described organizing different aspects of a task across branches: \textit{``Something like… day one this was the diet plan… what was the exercise routine… different kind of simultaneously, but two different branches.''} This ability to externalize parallel lines of thought supported more structured and efficient exploration.

Branching was also used to systematically explore variations of the same idea through prompt reuse. Participants iterated on prior inputs while preserving earlier results. For example, P10 noted, \textit{``I reused the previous prompts… to create different branches… what if I had one hour per day… or 20 minutes per day?''}, while P5 described applying the same prompt across conceptual variations: \textit{``The same prompt we used, but with a different… concept… polymorphism… virtualization… singleton pattern…''}. This pattern enabled controlled experimentation across alternatives without losing context. Participants further revisited earlier states and resumed exploration from previously abandoned paths. As P7 explained, \textit{``I went to the Canvas option… went back in conversation where I abandoned another idea, and then… picked up the conversation from there again.''} This iterative navigation allowed users to treat conversations as revisitable states rather than linear histories. However, P20 noted, \textit{``Switching between canvas and chat was confusing''}. This difficulty highlights a second adoption challenge distinct from branch
semantics. Although the synchronized views exposed the same underlying conversation, moving between linear and spatial representations required
participants to maintain orientation across two different interaction models. For some participants, the benefit of obtaining a structural overview, therefore, came with an additional coordination cost when returning to a focused
conversation.

Participants also noted limitations in comparing and integrating outcomes across branches. While the system supported divergence, synthesizing results remained challenging. As P3 suggested, \textit{``maybe merging the two ideas to see how would they look like''}. Overall, while branching enabled parallel exploration and iteration, it introduced new challenges around cross-branch comparison and integration.

\subsubsection{Reflection and Interpreting the Spatial Representation}

Participants reported that the system supported reflection and helped them better understand their own thinking processes. By externalizing conversation structure, the canvas enabled users to reason about their progress and decision-making. As P23 noted, \textit{``It helped me understand how to approach the task''}. Participants also used branching for implicit comparison, with P11 explaining, \textit{``You can compare outputs very effectively by switching branches''}. This suggests that the representation supports metacognitive engagement during task execution.

The spatial layout was frequently interpreted through familiar structural metaphors, particularly maps and organized overviews. P14 described the canvas as \textit{``like a map where I could see how the conversation developed''}, while P19 emphasized clarity and structure: \textit{``It was very structured and I was able to [get an] overview''}. Participants highlighted that the arrangement of branches made relationships between ideas easier to interpret. As P11 stated, \textit{``The sub-conversations are ordered perfectly… very overview-friendly''}. 

In addition, the representation supported fluid transitions between overview and detail, enabling users to navigate different levels of abstraction. P6 explained, \textit{``you can see from the overview, and then you go into details''}, and noted that branching allows ideas to be considered \textit{``side by side… otherwise you go up and down… but here… it helps you think in an organized way.''} 

Across these experiences, appropriation of the spatial representation was therefore uneven rather than universal. Some participants incorporated the canvas and branching into their workflows as mechanisms for organizing, revisiting, and comparing conversational paths, while others continued to rely on familiar linear interaction or encountered difficulties understanding branch context and coordinating between views. These differences suggest that adopting a non-linear conversational interface involved developing new interaction practices and mental models, rather than simply learning where interface features were located.

\subsection{Subjective Evaluation}
% \rifat{So both evaluations are subjective but I think this is quantitative so I will switch the title of this section.}
In this section we discuss participants' subjective experience of CanvasConvo using standardized usability and workload measures (SUS, NASA-TLX), along with Likert-scale ratings of agency, authorship, and system support. These measures capture how users perceived control, collaboration, and their ability to manage complex, multi-branch interactions.

\subsubsection{System Usability Score}
For the system usability score (SUS), CanvasConvo achieved a mean score of $78.9$ ($SD = 11.2$), with a median of $80.0$ and a range from $56.7$ to $94.4$
% (see \autoref{fig:SUS_field})
. This places the system well above the standard usability benchmark of 68, indicating strong overall usability. The majority of ratings fall within the ``good'' to ``excellent'' range, with multiple participants assigning scores above $85$, reflecting consistently positive perceptions of the system.

% \begin{figure}[ht!]
%     \centering
%     \includegraphics[width=\linewidth, trim={1cm 0.5cm 1cm 2.5cm},clip]{figures/SUS.pdf}
%     \caption[System Usability Scale (SUS) results from our field-study.]{SUS results from our field study (N=24).}
%     \label{fig:SUS_field}
%     \Description{This image shows the System Usability Score for the second version of our system.}
% \end{figure}

% \subsubsection{Mental Load}
\subsubsection{Perceived Workload (NASA-TLX)}
We analyzed raw NASA-TLX ratings for our tool to assess perceived workload. Overall, the results indicate low workload across all factors (see \autoref{tab:NASA-TLX} in Appendix). Participants reported particularly low levels of physical demand ($M = 1.50, SD = 1.10$) and annoyance ($M = 1.67, SD = 0.82$), while mental demand ($M = 2.83, SD = 1.55$) and effort ($M = 3.29, SD = 1.68$) were slightly higher but still in a low-to-moderate range. The overall NASA-TLX score ($M = 2.79, SD = 0.74$) further suggests that participants experienced the system as lightweight and manageable.
% \begin{table}[h!]
%     \caption{The raw NASA-TLX results (N=24).}
%     \begin{tabularx}{0.7\linewidth}{X d{2.2} d{2.2}}
%     \toprule
%        \textbf{Factor}  & \multicolumn{1}{c}{\textbf{M}}  & \multicolumn{1}{c}{\textbf{SD}}\\ 
%        \midrule
%        Mental Demand      & 2.83 & 1.55 \\
%        Physical Demand    & 1.50 & 1.10 \\
%        Temporal Demand    & 1.96 & 1.16 \\
%        Success            & 5.46 & 1.14 \\
%        Effort             & 3.29 & 1.68 \\
%        Annoy              & 1.67 & 0.82 \\
%        \midrule
%        Overall NASA-TLX (Sum)  & 16.71 & 4.44 \\ 
%        Overall NASA-TLX (Mean) & 2.79 & 0.74 \\ 
%     \bottomrule
%     \end{tabularx}
%     \Description{This table shows the NASA-TLX results for the evaluated tool (N=24).}
%     \label{tab:NASA-TLX}
% \end{table}

% \subsection{System Usability}
% \rifat{start with qualitative findings}

% \subsubsection{In general, others...}

\subsubsection{Authorship and Agency}

\begin{figure}[t]
    \centering
    \includegraphics[width=\linewidth]{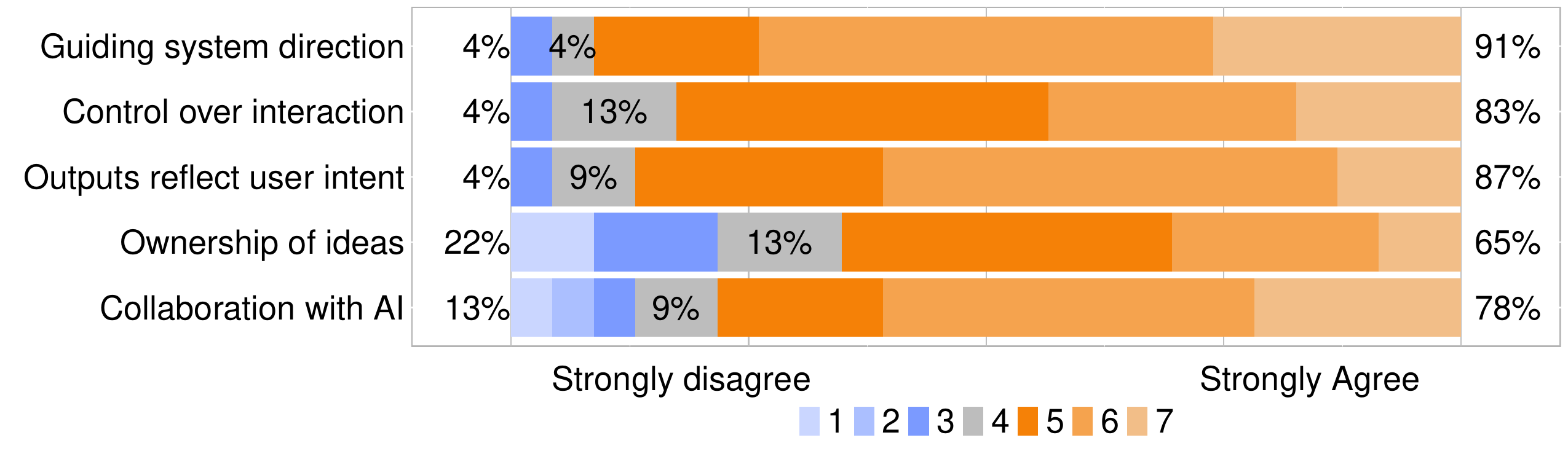}
    \caption{Participant ratings of perceived agency, authorship, and collaboration when using the system, measured on a 7-point Likert scale (N=24).} 
    % Responses are predominantly positive, with most participants agreeing they could guide the system and that outputs reflected their intentions.}
    \label{fig:authorship}
    \Description{This image shows }
\end{figure}

Participants reported a generally strong sense of agency and alignment when interacting with the tool, see \autoref{fig:authorship}. Most participants agreed that they could easily guide the system (91\%) and felt in control of the conversation and its outcomes (83\%). Similarly, a large majority indicated that the outputs reflected their own ideas and intentions (87\%), suggesting that the system supports user-driven expression rather than overriding it. While perceptions of ownership were slightly more mixed, a majority still felt that the ideas developed during the interaction belonged to them (65\%). Participants largely experienced the interaction as collaborative (78\%), indicating that the tool was seen not just as a passive assistant but as an active partner in the process.

\subsubsection{Perceived System Support}
\begin{figure}[t]
    \centering
    \includegraphics[width=\linewidth]{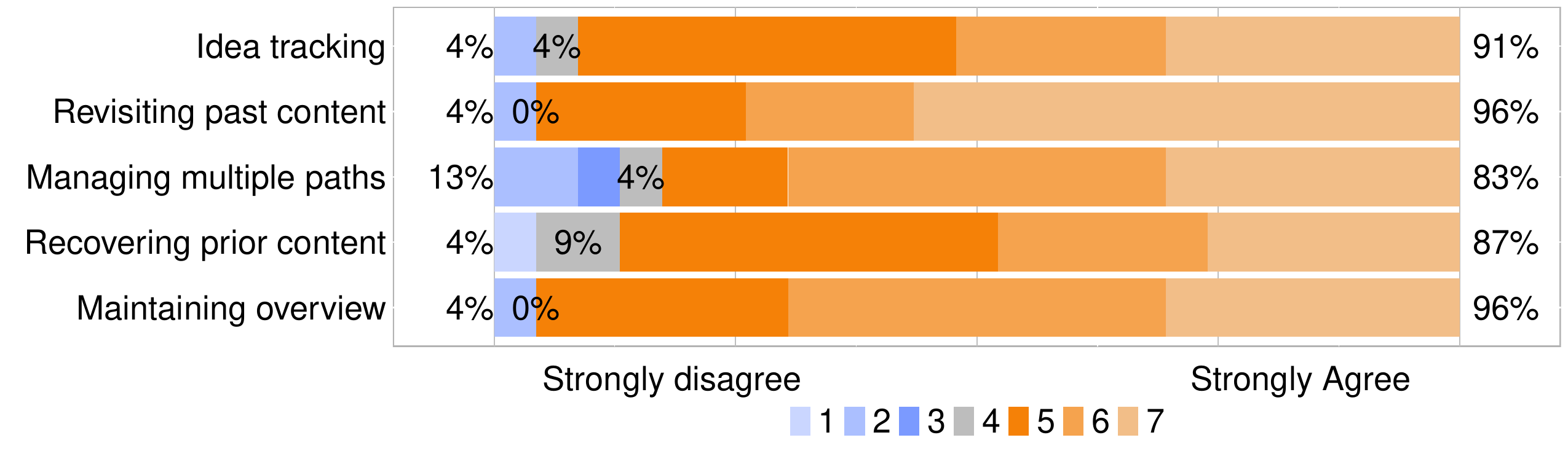}
    \caption{Participant ratings of the system's support for overview and context management across interaction, measured on a Likert scale (N=24).} 
    % Responses are predominantly positive, with most ratings falling on the agreement side of the scale. Colors encode response categories from strongly disagree (left) to strongly agree (right).}
    \label{fig:system}
    \Description{This image shows }
\end{figure}

Participants also reported strong support for managing and navigating longer, more complex interactions (see \autoref{fig:system}). A large majority agreed that the system helped them keep track of ideas across extended conversations (91\%) and maintain an overview of their work (96\%). Nearly all participants found it easy to revisit earlier parts of the conversation (96\%), indicating effective support for accessing prior context. Similarly, most participants reported that it was easy to recover previous prompts or results when needed (87\%). While slightly more varied, responses still showed that participants felt supported in managing multiple directions or ideas within a conversation (83\%). 

Overall, participants rated the system positively for organization, navigation, and continuity in multi-step interactions. These aggregate ratings should, however, be interpreted alongside the qualitative findings, which show that participants differed in how readily they incorporated the spatial representation into their workflows and in how clearly they understood its interaction semantics.

% Overall, these results suggest that the system effectively addresses challenges of organization, navigation, and continuity in multi-step LLM interactions.

% editing this figure for space

\subsection{Objective Telemetry}
To complement participants' self-reported experiences, we analyzed interaction logs collected through a custom instrumentation layer during the field study. Across 24 participants, the system recorded 165 distinct conversations and 1,225 messages. On average, participants generated 51.04 messages and 6.88 conversations, indicating sustained engagement with the system over multiple sessions. Summary statistics are presented in \autoref{tab:telemetry-field} in the Appendix.

\subsubsection{Feature Engagement and Interaction Distribution.}
Interaction activity was unevenly distributed across system components. Chat-based interactions accounted for the largest proportion of events (47.2\%), followed by canvas interactions (28.3\%) and sidebar navigation (11.4\%). In addition, navigation behaviors such as branch revisits and timeline interactions were frequently observed, with 93 branch navigation events and 132 timeline clicks, highlighting active use of the system's mechanisms for revisiting prior context. Prompt templates were used 18 times in total, and 5 participants created templates.

The distribution, therefore, does not indicate a shift from chat-centered to canvas-centered interaction during the deployment. Instead, participants continued to perform most recorded interactions through the familiar chat interface while selectively engaging with the spatial representation and its navigation mechanisms. This pattern is consistent with the qualitative findings in \autoref{results:explorationbranching}, and \autoref{results:spatialStructure}, where participants described the canvas as particularly useful for obtaining an overview, revisiting prior states, and managing alternative directions, but not necessarily as their primary surface for conversational interaction.

Prompt templates showed substantially narrower adoption: only five participants created templates during the deployment. This suggests that reusable prompting was not a central part of most participants' workflows during the study, despite being available as a supporting mechanism.

\subsubsection{Temporal Patterns of Use.}
% A comparison of session durations reveals a clear asymmetry between chat and canvas usage. Canvas sessions lasted substantially longer (M = 34.1 minutes) than chat sessions (M = 13.0 minutes). 
% \done{sentence below: This can also happen because of the lack of familiarity with the interface. Maybe they accomplish the same goals faster with the linear flow and the chat-based one and they take longer to do the same with the branch one. Is there any way to disentangle this with the current data?} Aligning with the findings in \autoref{results:explorationbranching}, this suggests that participants used the chat interface primarily for rapid input and iteration, whereas the canvas supported longer periods of interaction, such as reviewing, organizing, and reflecting on prior content. This distinction indicates complementary roles of the two interfaces.

A comparison of recorded view durations shows that canvas-view events had a higher mean duration ($M=34.1$ minutes) than chat-view events ($M=13.0$ minutes). This difference should be interpreted cautiously. The telemetry records wall-clock time between entering and leaving a view and does not distinguish active interaction from reading, inactivity, or time spent in a background tab. Moreover, participants may have selected the canvas for more complex tasks, and lower familiarity with the spatial interface may itself have increased interaction time. The current data therefore cannot determine whether longer canvas durations reflect more sustained organization and reflection, greater task complexity, lower interface familiarity, or a combination of these factors.

In conjunction with the qualitative findings in \autoref{results:explorationbranching}, the duration pattern is consistent with participants using chat for relatively rapid input and iteration and the canvas for longer episodes of reviewing and organizing prior content. However, the telemetry alone does not establish these functional roles. A controlled within-participant study using equivalent tasks, active-time measurements, and task-level covariates would be needed to disentangle interface effects from differences in familiarity and task complexity.

\subsubsection{Branching and Exploratory Behavior.}
% Participants actively engaged with branching as a mechanism for exploring alternative conversational trajectories. In total, 79 branches were created (M = 3.29 per participant), including 15 sub-branches and a maximum nesting depth of two levels. Additionally, branch navigation events (N = 93; M = 3.88) slightly exceeded branch creation, \rifat{This could be confirmation bias because this is almost one on one, so I will not say they frequently revisit the branches.} indicating that participants frequently revisited existing branches. These patterns support our qualitative findings in \autoref{results:explorationbranching}, suggesting that users appropriated branching to structure conversations as navigable trees rather than linear sequences.

Participants engaged with branching as a mechanism for creating alternative conversational trajectories. In total, 79 branches were created ($M=3.29$ per participant), including 15 sub-branches and a maximum nesting depth of two levels. The telemetry also recorded 93 branch-open events ($M=3.88$ per participant), indicating that participants navigated to existing branches in addition to creating them.

However, because the number of branch-open events only modestly exceeded the number of branch-creation events, these aggregate counts do not establish that branches were frequently revisited. The data also do not distinguish an immediate opening of a newly created branch from a later return to an earlier branch. We therefore interpret these events more conservatively as evidence that participants both created and navigated among branching conversation paths. Together with the qualitative findings in \autoref{results:explorationbranching}, this suggests that at least some participants used branching to structure alternatives as navigable conversational trees rather than solely as linear sequences.

\subsubsection{Interaction with the Spatial Canvas.}
Telemetry from the canvas interface highlights active engagement with spatial navigation features. Zoom interactions were the most frequent action ($N = 881$; $M = 36.71$), followed by node selections ($N = 246; M = 10.25$) and search usage ($N = 94; M = 3.92$). The high frequency of zoom interactions shows that participants frequently
changed the level of detail displayed on the canvas. Together with participants'
descriptions of moving between overview and detailed inspection,
this pattern is consistent with the use of the canvas for navigating across
levels of conversational detail. This aligns with the results discussed in \autoref{results:spatialStructure}. Temporal navigation was also actively used. Participants clicked on timeline entries 132 times ($M = 5.50$), directly navigating to earlier points in the conversation. This provides behavioral evidence supporting participants' reported ease of revisiting prior context (\autoref{results:explorationbranching}).

% \done{for the pargraph below: I will say it will be good to revisit the claims in this section because many of them don't clearly seem to be supported by the data. Maybe they require further analysis or a causal relationship analysis.}

% Overall, the telemetry data reveals a complementary usage pattern across system components. Chat interactions primarily supported content generation and task progression, while the canvas and timeline were used for orientation, navigation, and reflection. The combination of longer canvas sessions, frequent navigation actions, and active branching behavior suggests that participants engaged with the system as a non-linear conversational workspace rather than a traditional linear chat interface.

Overall, the telemetry data complement our qualitative findings by revealing distinct interaction patterns across the interface components. Chat interactions were more frequent, while canvas sessions tended to be longer and were accompanied by navigation actions such as branching, node selection, and timeline use. Although these behavioral traces do not by themselves establish how participants used each interface, they are consistent with interview accounts describing chat as supporting rapid conversational interaction and the canvas as facilitating organization and exploration during longer-running tasks.

\section{Discussion \& Future Work}
In this work, we investigated how spatial and non-linear conversational interfaces can support LLM-based workflows. Specifically, we examined (RQ1) how users manage multiple ideas and alternative directions in traditional LLM chat interfaces, (RQ2) how non-linear structures influence reuse and navigation of conversational artifacts, and (RQ3) how users adopt and appropriate spatial conversational interfaces in longer-running tasks. 

% \done{The evaluation surfaced notable friction in the interface design that the current discussion treats as mere caveats rather than core findings to grapple with:

% 1. For instance, telemetry shows that traditional chat interactions heavily dominated over canvas interactions, and multiple participants reported confusion regarding branching semantics (i.e., not knowing what context a branch retains)

% % 2. Additionally, some features in the wide list of tools (e.g., the writing toolbar) do not feel clearly grounded in the core problem
% }

% \rifat{refer to the RQs down below}

% Our findings highlight both the potential of structuring conversations as manipulable artifacts and the tensions introduced by increased interaction complexity and learning overhead.

\paragraph{Limitations}
One limitation of our study is the duration and scope of the field deployment. While the five-day study allowed us to observe evolving workflows and repeated interactions, longer-term use may reveal different patterns of adoption. %, especially as users develop stronger mental models of branching and spatial organization. 
Additionally, our participant pool consisted largely of experienced LLM users, which may limit generalizability to novice populations. Future work should investigate how such systems are adopted over extended periods and across different expertise levels. %, particularly focusing on onboarding and learning curves.
A further limitation concerns variability in participants' engagement and task characteristics. Participants worked on self-selected activities that differed considerably in duration, complexity, and information density, resulting in different cognitive and sensemaking demands across individuals. Finally, subjective measures such as NASA-TLX should be interpreted cautiously, as observed workload differences may partially reflect variation in underlying tasks rather than interface effects alone. Future studies could employ more controlled task sets alongside ecological deployments to better isolate the effects of interaction design. 

% Finally, the five-day deployment captures early appropriation rather than
% long-term stabilization of interaction practices; participants who continued to rely on familiar chat behaviors during the study might adopt the spatial
% representation differently with extended use, but our data cannot establish whether or how such practices would change.

% \done{another limitation: Each participant could potentially have different levels of engagement, different amount of information they worked with, which would placie different sense-making demands, etc - these are confounds for tests like the NASA TLX}

\paragraph{Spatial Representations Complement Rather Than Replace Linear Chat}
Our findings suggest that the transition from linear to spatial conversational interaction is better understood as appropriation than replacement. Participants continued to rely heavily on the familiar chat interface while selectively incorporating the canvas when they needed to revisit prior states, obtain an overview, or manage alternative directions. This division of labor is reflected both in participants' accounts and in the interaction telemetry, where chat interactions remained more frequent than canvas interactions. Importantly, adoption of the spatial representation was uneven. P19's continued reliance on scrolling illustrates how established interaction habits can persist even when alternative navigation mechanisms are available. P20's difficulty switching between canvas and chat highlights the coordination cost introduced by maintaining synchronized linear and spatial representations, while P23's uncertainty about inherited branch context shows that users must also develop an accurate mental model of non-linear conversation semantics. Together, these findings suggest that designing non-linear LLM interfaces requires supporting transitions between representations and communicating conversational state, not simply exposing additional structure.

\paragraph{Externalizing Conversations as Navigable Structures} 
% \textcolor{red}{(RQ2)}
Our results show that representing conversations as spatial structures changes how users interact with LLMs. Instead of treating conversations as long transcripts, participants engaged with them as persistent, navigable artifacts, often describing the system as a ``map'' or ``tree'' of ideas. This aligns with prior work emphasizing the importance of external representations for sensemaking and exploration~\cite{Jiang_Rayan_Dow_Xia_2023, promptcanvas}. Unlike such prior approaches that visualize single LLM responses \cite{Jiang_Rayan_Dow_Xia_2023, Zhang_2023}, CanvasConvo structures the entire interaction as a branching history, enabling parallel exploration and better workflow awareness. Features like spatial layout, timeline navigation, and semantic zoom supported smooth transitions between overview and detail \cite{shneiderman1983direct}, as also reflected in the interaction logs. Additionally, search, tagging, and automatic summaries helped users navigate and retrieve information in larger conversations.

\paragraph{Managing Cognitive Demands}
%\rifat{Generative AI tools often increase metacognitive load by requiring users to manage complex prompts \cite{metacognition}}
People's use of our system and their reflections in comparison to their usual tools suggest that the linear chat paradigm is effective for short, goal-directed tasks but breaks down in exploratory and multi-step workflows. Concretely, for their usual LLM-based tools, they described difficulties navigating verbose outputs, maintaining context, and comparing alternatives, often resorting to external tools or fragmented strategies such as copying prompts or splitting tasks across chats. These observations align with prior work showing that LLM interaction introduces significant metacognitive demands, particularly in monitoring progress and evaluating outputs~\cite{FERNANDES2026108779, metacognition}. 
% \rifat{link to our results}
Based on participants' experiences in our study, by externalizing conversational structure, CanvasConvo %reduces 
is perceived as reducing reliance on memory and scrolling. %, enabling users to offload cognitive effort into the interface. 
However, these new representations come with interaction demands. Future designs could explore further combinations, transitions, and adaptations of linear and non-linear conversation UIs. % suggesting that future systems must carefully balance structural richness with cognitive simplicity.

% \paragraph{From Prompting to Structured Interaction}
% A key implication of our findings is that prompting alone is insufficient as the primary interaction paradigm, particularly for \textcolor{red}{non-expert users}. Consistent with prior work~\cite{johny}, participants struggled to reliably express and refine intent through prompts, often relying on trial-and-error strategies. 

% CanvasConvo addresses this by introducing explicit mechanisms for structuring intent, such as context panels and reusable prompt templates. Interestingly, participants reported increased perceptions of control and agency even when such features were not actively used, suggesting that the \textit{availability} of structured interaction mechanisms can shape user experience. This points to an important design implication: interfaces should not only support control but also make it visible and accessible.

\paragraph{Branching as a Mechanism for Exploration}
%Branching appeared as a key mechanism for enabling exploration. 
Our branching feature was accepted and leveraged for exploring \textit{what-if} scenarios. Unlike linear chats, it lets users explore alternatives without restarting, while preserving context. Participants used it to test ideas, compare approaches, and separate lines of reasoning, lowering the barrier to exploration \cite{suh2024luminate}. Branching also supported users' sense of control, as branches were seen as independent spaces. 

To compare results across branches, participants developed their own strategies, such as adopting reusable prompt templates to summarize and transfer insights. %This shows how users adapted the system for synthesis. 
Since the conversation is structured as a tree, it supported divergence but not merging. Future designs could 
% introduce dedicated features for combining ideas across branches.
address this gap by introducing mechanisms for merging and reconciling branches. For example, such designs could draw inspiration from merging in version control systems such as Git. %thereby enabling users to not only explore alternatives, but also systematically combine them into coherent outcomes.

At the same time, our findings suggest that branching requires clearer interaction semantics than current implementations provide. While participants appreciated the ability to isolate ideas, several were uncertain about exactly what conversational state was preserved across branches. This ambiguity occasionally reduced confidence in the system despite branching operating correctly. These observations suggest that future systems should make context inheritance explicit—for example through visual indicators of shared history, inherited prompts, or branch-specific summaries—to better align users' mental models with system behavior.

\paragraph{Supporting Multi-Step and Long-Running Workflows} 
% \textcolor{red}{(RQ3)}
In longer-running own tasks, participants viewed CanvasConvo as a workspace rather than a chat tool, as it supported parallel exploration, iterative refinement, and revisiting prior states (with timeline navigation). This is consistent with work showing that externalizing cognitive work into manipulable representations supports more effective reasoning across complex, multi-step tasks \cite{kirsh_thinking_2010}.
Working with CanvasConvo thus supported planning, ideation, and problem-solving processes. Telemetry data further supports this distinction. Interestingly, participants naturally developed a division of labor between interaction modes. Linear chat supported rapid iteration and prompt refinement, whereas the spatial workspace was primarily used for organization, revisiting previous work, and managing multiple concurrent directions. 

% Rather than replacing conversational interfaces, our findings suggest that spatial workspaces complement them by supporting forms of interaction that become increasingly important as conversational tasks grow in duration and complexity. However, this richer interaction adds demands related to navigation across views and understanding system behavior (e.g., branch context). While the views were linked and supported by features such as our timeline, future work could explore their combination and integration further. 

% indicate that richer interaction requires clearer mental models and guidance.

% \paragraph{Trade-offs Between Expressiveness and Usability}
% Our findings reveal a fundamental trade-off between expressive power and usability. While many participants valued the flexibility and control offered by CanvasConvo, others perceived the interaction as complex or effortful, particularly when compared to familiar linear chat interfaces. Adoption was strongly influenced by prior habits and mental models, with some users reverting to linear interaction patterns despite the availability of non-linear features. This suggests that spatial conversational interfaces may be most beneficial for complex, exploratory tasks, while simpler interfaces may remain preferable for quick, focused interactions. Designing adaptive systems that can fluidly transition between linear and non-linear modes may help address this tension.

\paragraph{Authorship, Agency, and Control in Non-Linear Interaction}
%\rifat{prior systems such as Wordcraft~\cite{wordcraft} and professional-writer studies~\cite{ippolitoCreativeWritingAIPowered2022}, issues of authorship, disclosure, and cognitive ownership arise when human and AI collaborate.}

Participants reported a strong sense of agency when interacting with CanvasConvo.
Prior work on human-AI co-writing has shown that close integration of AI raises questions of authorship, disclosure, and cognitive ownership \cite{wordcraft, ippolitoCreativeWritingAIPowered2022}. In our study, the introduction of branching supported a sense of ownership by making the model's context visible and controllable.
%, particularly in their ability to guide the system, control conversational direction, and shape outcomes. The 
%introduction of branching supported this sense of control 
This allowed users to isolate ideas, explore alternatives independently, and avoid unintended carryover between interaction paths. Related, participants' perceived that outputs largely reflected their own intentions. %, suggesting that structuring interaction as user-driven exploration preserves authorship even in AI-assisted workflows. 
The system was frequently perceived as a collaborative partner rather than a passive tool, indicating a shift toward co-creative interaction. These findings suggest that non-linear interfaces can strengthen perceived agency by identifying user decisions and interaction paths. %while the user can identify their decisions and interaction path?

\section{Conclusion}
We presented CanvasConvo, a conversational interface that transforms linear LLM interaction into a non-linear, spatial workspace through branching, a canvas, and timeline-based navigation. In a 5-day field study with 24 participants, we observed how users structured multi-step workflows, explored alternatives in parallel, and revisited prior interactions within this space. Participants used branching to separate ideas and explore directions, while the canvas, timeline, and supporting features (e.g., summaries, tags, reusable prompts) helped maintain an overview and reuse conversational content. These findings show how conversation-level spatial representations can support exploration, navigation, and sensemaking in LLM-based work, while also revealing that their adoption is neither immediate nor uniform. Participants selectively combined spatial interaction with familiar linear chat practices, and the added structure introduced challenges in navigation, switching between views, and understanding conversational context across branches. Overall, our results suggest that future LLM interfaces should move beyond linear chat toward interactive workspaces that support both exploration and synthesis, while balancing structure with usability.

% We presented CanvasConvo, a conversational interface concept that transforms linear LLM chat into a non-linear, spatially organized workflow. By introducing branching, temporal navigation, and a spatial canvas, the system enables users to explore alternatives, revisit prior states, and structure conversations as persistent artifacts. Through a 5-day field study with 24 participants, we showed that while linear chat supports simple tasks, it breaks down for multi-step and exploratory workflows, leading users to adopt fragmented strategies.  \daniel{The study did not really show that... this sentence above sounds like people used their ususal tools over these 5 days.} In contrast, CanvasConvo supports parallel exploration, improves navigation and externalizes conversational structure. However, these benefits come with trade-offs: Increased expressiveness introduces challenges in usability, mental models, and adoption. Our findings highlight the need to move beyond linear chat toward interfaces that balance structure and flexibility. We argue that future LLM interfaces should treat conversations not as logs, but as manipulable, explorable workspaces.

\begin{acks}
% Funded by Elitenetzwerk Bayern. 
% This project is funded by the Deutsche Forschungsgemeinschaft (DFG, German Research Foundation) -- 525037874. 
Funded by Elitenetzwerk Bayern and the Deutsche Forschungsgemeinschaft (DFG, German Research Foundation) -- 525037874.
\end{acks}

%%
%% The next two lines define the bibliography style to be used, and
%% the bibliography file.
\bibliographystyle{ACM-Reference-Format}
\bibliography{sample-base}

% \newpage
%%
%% If your work has an appendix, this is the place to put it.
\appendix
\onecolumn

% \begin{table*}[ht!]
% \centering
% \caption{Objective telemetry data from the main study for \textbf{CanvasConvo}. Values represent per-participant averages.}
% \label{tab:telemetry-field}
% \begin{tabular}{lc}
% \toprule
% \textbf{Telemetry} & \textbf{Value} \\
% \midrule

% % Total events (per participant)         & 290.29 \\
% Messages              & 51.04  \\
% Conversations         & 6.88   \\ \\

% \textbf{Interaction distribution (\%)} &         \\
% Chat interactions                     & 47.2\%  \\
% Canvas interactions                   & 28.3\%  \\
% Sidebar navigation                    & 11.4\%  \\ \\

% \textbf{Session duration (minutes)}    &         \\
% Canvas sessions                       & 34.10   \\
% Chat sessions                         & 13.00   \\ \\

% \textbf{Branching behavior}            &         \\
% Branches created     & 3.29    \\
% Branch navigation events              & 3.88    \\ \\

% \textbf{Canvas interaction events}     &         \\
% Zoom level changes                    & 36.71   \\
% Node clicks                           & 10.25   \\
% Canvas search usage                   & 3.92    \\
% Timeline navigation clicks            & 5.21    \\

% \bottomrule
% \end{tabular}
% \end{table*}

\section{Methodology}

\subsection{Participants}
\begin{table}[htbp]
\centering
\caption{Participant Overview (pre-task questionnaire Survey): Demographics, professional background, familiarity with LLM tools, usage frequency, and primary tasks.}
\label{tab:participant-overview2}
\footnotesize
\begin{tabular}{l l l p{2.2cm} p{0.5cm} p{1.5cm} p{1.9cm} p{3.6cm}}
\toprule
Nr. & Gender & Prof. Writer & Job & Age & Familiarity & Frequency & LLM Tasks Used \\
\midrule
P1 & Female & No & PhD student & 27 & Very familiar & Daily & Writing/editing; coding \\
P2 & Female & No & PhD Student & 28 & Very familiar & Multiple/day & Writing/editing; coding; data analysis; planning; brainstorming \\
P3 & Male & No & Software Engineer & 32 & Moderately familiar & 3–5/week & Writing/editing; coding; data analysis; learning \\
P4 & Male & No & Accountant & 29 & Moderately familiar & 3–5/week & Writing/editing; brainstorming; learning \\
P5 & Male & Yes & Data Engineer & 34 & Expert & Multiple/day & Writing/editing; coding; data analysis; brainstorming; learning \\
P6 & Female & No & Sales & 29 & Moderately familiar & 3–5/week & Writing/editing; brainstorming; learning \\
P7 & Female & Yes & Research Communication Specialist & 29 & Very familiar & 1–2/week & Writing/editing; planning; brainstorming; learning \\
P8 & Non-binary & No & Student & 28 & Very familiar & 3–5/week & Writing/editing; coding; data analysis; learning \\
P9 & Male & Yes & Researcher & 33 & Expert & Daily & Writing/editing; coding; brainstorming \\
P10 & Male & No & Data Science Specialist & 31 & Very familiar & Multiple/day & Writing/editing; coding; data analysis; planning; brainstorming; creative writing \\
P11 & Female & No & Master's Student & 26 & Very familiar & Daily & Writing/editing; coding; brainstorming; learning \\
P12 & Male & No & Visual Designer & 27 & Very familiar & Daily & Writing/editing; brainstorming \\
P13 & Male & No & Unemployed & 35 & Very familiar & Daily & Coding; brainstorming; learning \\
P14 & Male & No & Software Engineer & 28 & Very familiar & 3–5/week & Coding; data analysis; planning; brainstorming; learning \\
P15 & Male & No & CS Master Student & 28 & Very familiar & Multiple/day & Writing/editing; coding; data analysis; brainstorming; learning \\
P16 & Male & No & Software Developer & 29 & Very familiar & Multiple/day & Coding; data analysis; brainstorming; learning \\
P17 & Male & No & Intern (Full-Stack) & 21 & Very familiar & Daily & Writing/editing; coding; data analysis; planning; brainstorming; learning \\
P18 & Male & No & Student & 24 & Very familiar & Daily & Writing/editing; coding; data analysis; planning; brainstorming; learning \\
P19 & Male & No & Mechanical Eng. Intern & 24 & Very familiar & Multiple/day & Writing/editing; coding; data analysis; planning; brainstorming; learning \\
P20 & Male & No & Student & 25 & Very familiar & Multiple/day & Writing/editing; coding; data analysis; planning; brainstorming; learning \\
P21 & Male & No & CS Master Student & 21 & Expert & Multiple/day & Writing/editing; coding; data analysis; planning; brainstorming; learning; creative writing \\
P22 & Male & No & Researcher & 23 & Very familiar & 3–5/week & Writing/editing; coding \\
P23 & Male & No & GenAI Researcher & 54 & Expert & Multiple/day & Writing/editing; data analysis; planning; brainstorming \\
P24 & Male & No & Working Student (PM/Dev) & 22 & Very familiar & Multiple/day & Writing/editing; coding; data analysis; planning \\
\bottomrule
\end{tabular}
\end{table}

% \clearpage
% \newpage

\section{Results}

\subsection{NASA-TLX}
\begin{table}[h!]
    \caption{The raw NASA-TLX results (N=24).}
    \begin{tabularx}{0.4\linewidth}{X d{2.2} d{2.2}}
    \toprule
       \textbf{Factor}  & \multicolumn{1}{c}{\textbf{M}}  & \multicolumn{1}{c}{\textbf{SD}}\\ 
       \midrule
       Mental Demand      & 2.83 & 1.55 \\
       Physical Demand    & 1.50 & 1.10 \\
       Temporal Demand    & 1.96 & 1.16 \\
       Success            & 5.46 & 1.14 \\
       Effort             & 3.29 & 1.68 \\
       Annoy              & 1.67 & 0.82 \\
       \midrule
       Overall NASA-TLX (Sum)  & 16.71 & 4.44 \\ 
       Overall NASA-TLX (Mean) & 2.79 & 0.74 \\ 
    \bottomrule
    \end{tabularx}
    \Description{This table shows the NASA-TLX results for the evaluated tool (N=24).}
    \label{tab:NASA-TLX}
\end{table}

\subsection{Objective Telemetry}
\clearpage
\begin{table}[htbp]
\centering
\caption{Objective telemetry data from the main study for \textbf{CanvasConvo}. Values represent per-participant averages.}
\label{tab:telemetry-field}
\begin{tabular}{lc}
\toprule
\textbf{Telemetry} & \textbf{Value} \\
\midrule

Messages              & 51.04  \\
Conversations         & 6.88   \\ \\

\textbf{Interaction distribution (\%)} &         \\
Chat interactions                     & 47.2\%  \\
Canvas interactions                   & 28.3\%  \\
Sidebar navigation                    & 11.4\%  \\ \\

\textbf{Session duration (minutes)}    &         \\
Canvas sessions                       & 34.10   \\
Chat sessions                         & 13.00   \\ \\

\textbf{Branching behavior}            &         \\
Branches created                      & 3.29    \\
Branch navigation events              & 3.88    \\ \\

\textbf{Navigation events} & \\
Timeline navigation         & 5.50 \\ \\

\textbf{Canvas interaction events}     &         \\
Zoom level changes                    & 36.71   \\
Node clicks                           & 10.25   \\
Canvas search usage                   & 3.92    \\
Timeline navigation clicks            & 5.50    \\

\bottomrule
\end{tabular}
\end{table}

% \subsection{Part Two}

\end{document}